\RequirePackage{fix-cm}
\documentclass[twocolumn,epjc3]{svjour3}
\smartqed  
\RequirePackage{graphicx}
\RequirePackage{amsmath,amssymb,amsfonts}
\RequirePackage{multirow}
\RequirePackage{booktabs}
\RequirePackage{stfloats}
\RequirePackage{makecell}
\RequirePackage{hyphenat}
\RequirePackage{xurl}

\RequirePackage[mathlines]{lineno}

\journalname{Eur. Phys. J. C}

\begin{document}

\title{Prospects for precision Higgs boson measurements\\ at a 10 TeV muon collider
}

\titlerunning{Precision Higgs boson measurements at a 10 TeV muon collider}

\author{Paolo Andreetto\thanksref{addr1}
        \and
        Massimo Casarsa\thanksref{e1,addr2}
        \and
        Luca Castelli\thanksref{addr2}
        \and
        Alessio Gianelle\thanksref{addr1}
        \and
        Donatella Lucchesi\thanksref{addr1,addr3}
        \and
        Leonardo Palombini\thanksref{addr1,addr3}
        \and
        Imran Raghib\thanksref{addr4}
        \and
        Lorenzo Sestini\thanksref{addr4}
        \and
        Davide Zuliani\thanksref{addr1,addr3}
}

\authorrunning{P. Andreetto et al.}

\thankstext{e1}{e-mail: massimo.casarsa@ts.infn.it (corresponding author)}

\institute{INFN Sezione di Padova, Padova, Italy \label{addr1}
           \and
           INFN Sezione di Trieste, Trieste, Italy \label{addr2}
           \and
           Universit\`a di Padova, Padova, Italy \label{addr3}
           \and
           INFN Sezione di Firenze, Firenze, Italy \label{addr4}
}

\date{}

\maketitle

\begin{abstract}
A 10~TeV muon collider offers a powerful environment for precision studies of the Higgs sector through the large vector-boson-fusion production rates. This review presents projected sensitivities to single- and double-Higgs production cross sections, the Higgs boson mass, and the trilinear Higgs self-coupling, obtained from detailed simulations of the MUSIC detector including machine-induced background, assuming a baseline configuration of two experiments, each collecting an integrated luminosity of 10~ab$^{-1}$. The $H\to b\bar{b}$ and $H\to WW^\ast$ production cross sections can be measured with statistical precisions of 0.18\% and 0.35\%, respectively, while precisions of 2.6\%, 4.2\%, and 6.9\% are expected for $H\to\gamma\gamma$, $H\to ZZ^\ast$, and $H\to\mu^+\mu^-$. Combining the $H\to b\bar{b}$, $H\to\gamma\gamma$, and $H\to\mu^+\mu^-$ channels yields an expected Higgs boson mass precision of about 19~MeV. Double-Higgs production can be measured with statistical precisions of 4.2\% in the $H\!H\to b\bar{b}b\bar{b}$ channel and 14\% in $H\!H\to b\bar{b}WW^\ast$. Using the $H\!H\to b\bar{b}b\bar{b}$ channel, the trilinear Higgs self-coupling can be determined with an expected precision of about 5\%. These results demonstrate the potential of a high-energy muon collider for precision Higgs physics.
\end{abstract}

\section{Introduction}
\label{sec:intro}

The precise determination of the properties of the Higgs boson ($H$) is recognised as a central goal of high-energy physics in the coming decades,
as outlined in the 2026 update of the European Strategy for Particle Physics~\cite{esppu26} and in the 2023 report of the US Particle Physics
Project Prioritization Panel (P5)~\cite{P5-23}. This emphasis stems from the unique role of the Higgs field within the Standard Model (SM).

Indeed, the Higgs sector of the SM Lagrangian contains several key terms~\cite{PDG}. The Higgs boson couples to the $W$ and $Z$ bosons, with strengths proportional to the squares of their masses. Yukawa interactions between the Higgs field and the SM fermions generate the fermion masses after electroweak symmetry breaking (EWSB). The Higgs potential $V(H)$ contains the Higgs mass term together with the trilinear and quartic self-interaction terms.

Precise measurements of the Higgs boson couplings to elementary fermions and bosons therefore provide a powerful test of the mechanism of mass generation, and are sensitive to possible deviations induced by physics beyond the Standard Model (BSM). 
Percent-level precision in these measurements provides sensitivity to new physics at multi-TeV scales beyond the reach of current experiments~\cite{HiggsCouplings,HiggsStudies}.

Within the renormalisable SM, the shape of the Higgs potential is fully determined by two parameters, which can be expressed in terms of the Higgs boson mass and the Higgs self-coupling. If the SM is only an effective theory, many additional parameters may be present, introducing further degrees of freedom~\cite{peskinH}. A precise determination of the Higgs mass -- and in particular of the as-yet-unmeasured self-coupling -- is therefore crucial for stringent tests of the SM.

Among the future particle colliders proposed to address the fundamental questions raised by the Higgs sector, none is expected to answer all of them within a reasonable timescale~\cite{esppu26}.
A muon collider operating at a centre-of-mass energy of $\sqrt{s}=10$~TeV offers the opportunity to measure the Higgs couplings to fermions and gauge bosons with a precision comparable to that of other proposed facilities, while requiring substantially less running time and improving on the measurements anticipated at the High-Luminosity LHC (HL-LHC)~\cite{HL-LHC}.
More importantly, such a machine provides a unique opportunity to determine the Higgs boson self-coupling with the precision needed to probe the structure of the Higgs potential within a realistic experimental programme.

This paper presents the projected sensitivities for measurements of Higgs boson production cross sections, the Higgs boson mass, and the trilinear Higgs self-cou\-pling obtained with the MUSIC detector concept~\cite{music-paper}, optimised for operation at a $\sqrt{s}=10$~TeV muon collider. The studies assume an integrated luminosity of 10~ab$^{-1}$, corresponding to approximately five years of muon collider operation at the design luminosity, and focus on a set of benchmark channels, including the effects of both SM physics and machine-induced background.
The remainder of this paper is organised as follows.
Section~\ref{sec:hllhc} reviews the projected precision of Higgs boson measurements at the HL-LHC and discusses the physics opportunities offered by a high-energy muon collider. Section~\ref{sec:physics10tev} describes the Standard Model Higgs production mechanisms at a 10~TeV muon collider and the corresponding expected event yields. The Monte Carlo event generation, detector simulation, event reconstruction, and the common analysis methodology are presented in Sect.~\ref{sec:methodology}. The expected sensitivities for single-Higgs production, the Higgs boson mass, and double\hyp{Higgs} production are presented in Sects.~\ref{sec:measurements}, \ref{sec:mass}, and \ref{sec:selfcoupling}, respectively. Section~\ref{sec:xsec_summary} summarises the expected sensitivities for the benchmark Higgs production channels in the baseline two-experiment configuration, corresponding to a total integrated luminosity of 20~ab$^{-1}$. The expected precision on the trilinear Higgs boson self-coupling is presented in Sect.~\ref{sec:trilinear}. Finally, Sect.~\ref{sec:conclusions} summarises the main results and outlines the prospects for precision Higgs physics at a 10~TeV muon collider.

\section{The Higgs precision landscape after the HL-LHC}
\label{sec:hllhc}

Until the next major accelerator facility -- currently expected around 2050 -- the only experimental data on the Higgs boson will come from the LHC and its High-Luminosity upgrade, which is scheduled to conclude operations in the early 2040s.

\begin{figure}[!t]
    \centering
    \includegraphics[width=\columnwidth]{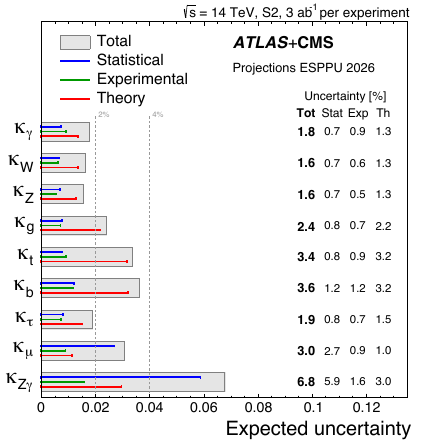}
    \caption{Combined ATLAS and CMS projected uncertainty on the Higgs coupling modifiers for $3~\mathrm{ab}^{-1}$ of $pp$ collisions per experiment in the $S2$ scenario, under SM assumptions. Figure taken from Ref.~\cite{HL-LHC_ESPPU}}
    \label{fig:ATLAS+CMS_HL-LHC-1}
\end{figure}

The projected precisions achievable by the end of the HL-LHC programme were recently updated by the ATLAS and CMS collaborations for the 2026 update of the European Strategy for Particle Physics~\cite{HL-LHC_ESPPU}. Two reference scenarios are considered for the systematic uncertainties. The $S2$ scenario assumes detector and analysis performance similar to that of previous projections, with reduced experimental systematic uncertainties and improved theoretical uncertainties. The $S3$ scenario further includes the advances in physics-object reconstruction and analysis procedures anticipated by the two experiments. Figure~\ref{fig:ATLAS+CMS_HL-LHC-1} shows the expected uncertainties on the Higgs coupling modifiers, defined as the ratio of the measured couplings to their SM expectations, for the $S2$ scenario and assuming the SM Higgs boson width. Most coupling uncertainties are projected to reach the few-percent level and are largely dominated by theoretical uncertainties, with statistical uncertainties dominating in the rare decay modes.

\begin{figure}[!t]
    \centering
    \includegraphics[width=\columnwidth]{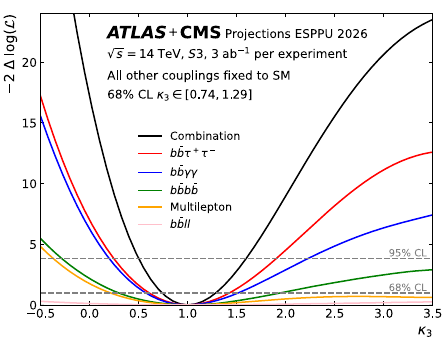}
    \caption{Expected ATLAS and CMS likelihood scans as a function of $\kappa_3$ for individual decay channels and their combination, with $3~\mathrm{ab}^{-1}$ of $pp$ collisions per experiment in the $S3$ scenario, assuming a true value of $\kappa_3 = 1$. Figure taken from Ref.~\cite{HL-LHC_ESPPU}}
    \label{fig:ATLAS+CMS_HL-LHC-2}
\end{figure}

The Higgs boson mass, $m_H$, is expected to be measured with a precision of about 21~MeV. Within the SM and under the $S2$ scenario assumptions, the Higgs boson total width can be constrained from off-shell production with an uncertainty of approximately 0.7~MeV.

The trilinear and quartic Higgs boson self-couplings are studied under the $S3$ scenario. They are conveniently expressed in terms of the coupling modifiers $\kappa_3 = \lambda_3 / \lambda_3^{\mathrm{SM}}$ and $\kappa_4 = \lambda_4 / \lambda_4^{\mathrm{SM}}$, where $\lambda_3$ and $\lambda_4$ denote the trilinear and quartic Higgs boson self-couplings, respectively, and $\lambda_{3,4}^{\mathrm{SM}}$ are their SM values.
The combined ATLAS and CMS projections for double-Higgs production with 3 ab$^{-1}$ per experiment yield an expected significance exceeding $7\sigma$.
Figure~\ref{fig:ATLAS+CMS_HL-LHC-2} shows the likelihood scans as a function of $\kappa_3$ for the decay channels considered in the measurement of the trilinear Higgs boson self-coupling. From the 
combined likelihood, the projected relative 68\%~CL interval is $[-26\%, +29\%]$ around the SM value.
For the European Strategy for Particle Physics Update, ATLAS and CMS have also presented a preliminary study of triple-Higgs production. 
Assuming the $S3$ scenario and building on LHC experience with six-$b$-jet final states, the projected 95\% CL upper limit on the cross section is approximately 86 times the SM prediction.

These expectations indicate that, while the HL-LHC will substantially refine measurements of Higgs couplings to gauge bosons and fermions, its sensitivity to the Higgs self-coupling will remain limited. 
Muon collisions at $\sqrt{s} = 10$~TeV can considerably extend the Higgs precision programme beyond the HL-LHC: the precision achievable on Higgs couplings to elementary fermions and gauge bosons is expected to be comparable to that of the HL-LHC, while the sensitivity to the Higgs self-coupling is projected to reach the percent level.

\section{Higgs production at a 10 TeV muon collider}
\label{sec:physics10tev}

The Higgs production mechanisms at multi-TeV muon colliders have been investigated in several theoretical studies, addressing different aspects of the production dynamics and the associated phenomenology. In this section, the benchmark SM production cross sections are taken from Ref.~\cite{Costantini:2020stv}, which provides a comprehensive and systematic study of the dominant Higgs production channels over a broad range of centre-of-mass energies. These benchmark values are used to illustrate the relative importance of the different production mechanisms and to provide a common reference for the discussion presented in this review.

\begin{figure*}[!t]
    \centering
    \includegraphics[width=0.325\textwidth]{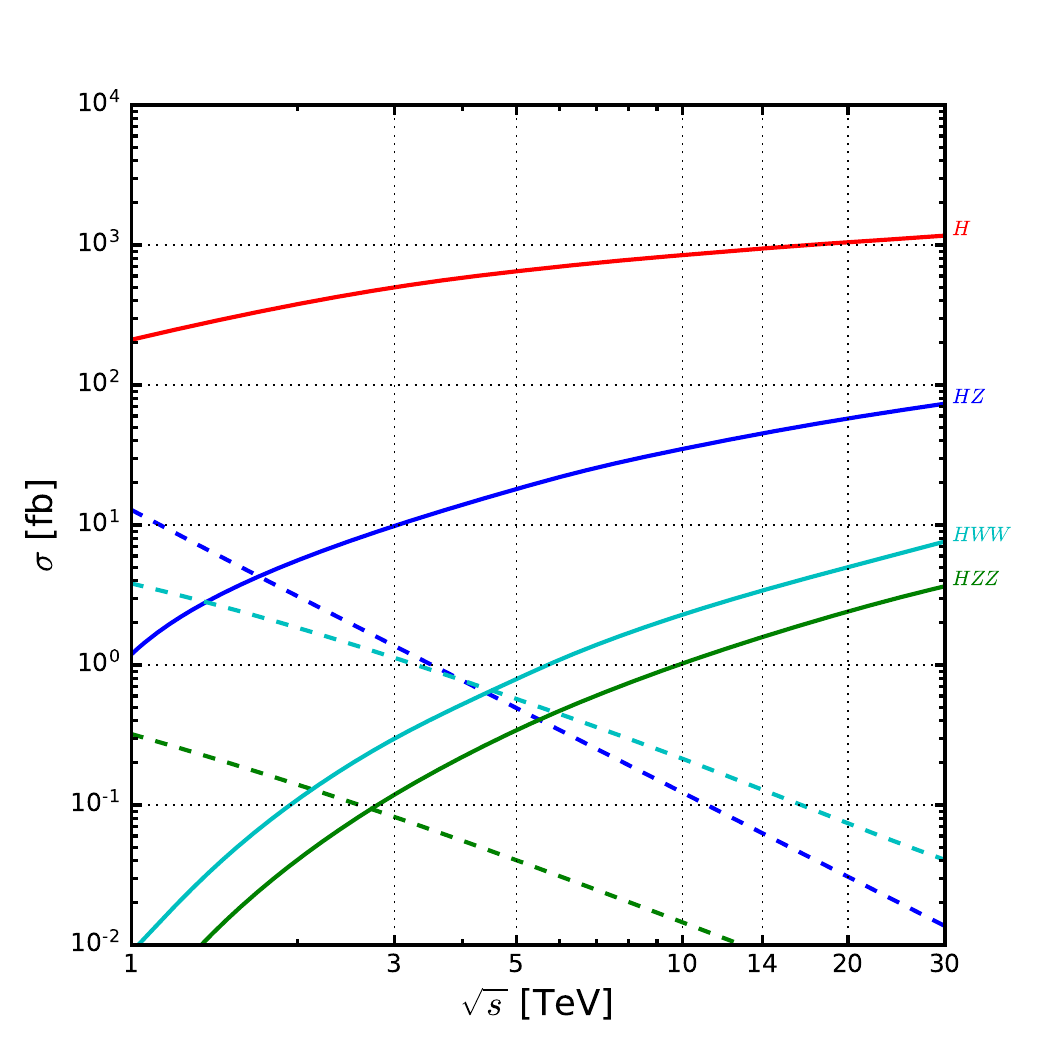}
    \includegraphics[width=0.325\textwidth]{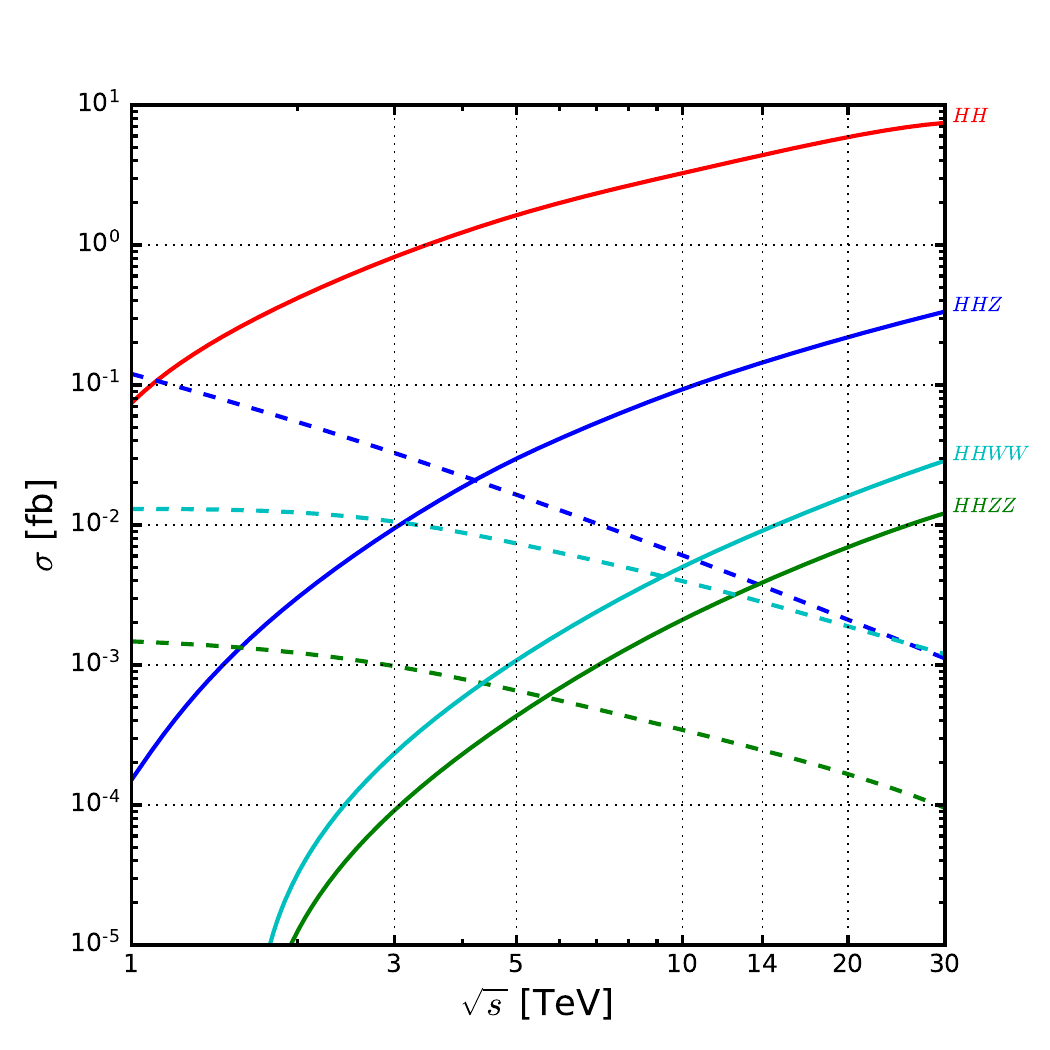}
    \includegraphics[width=0.325\textwidth]{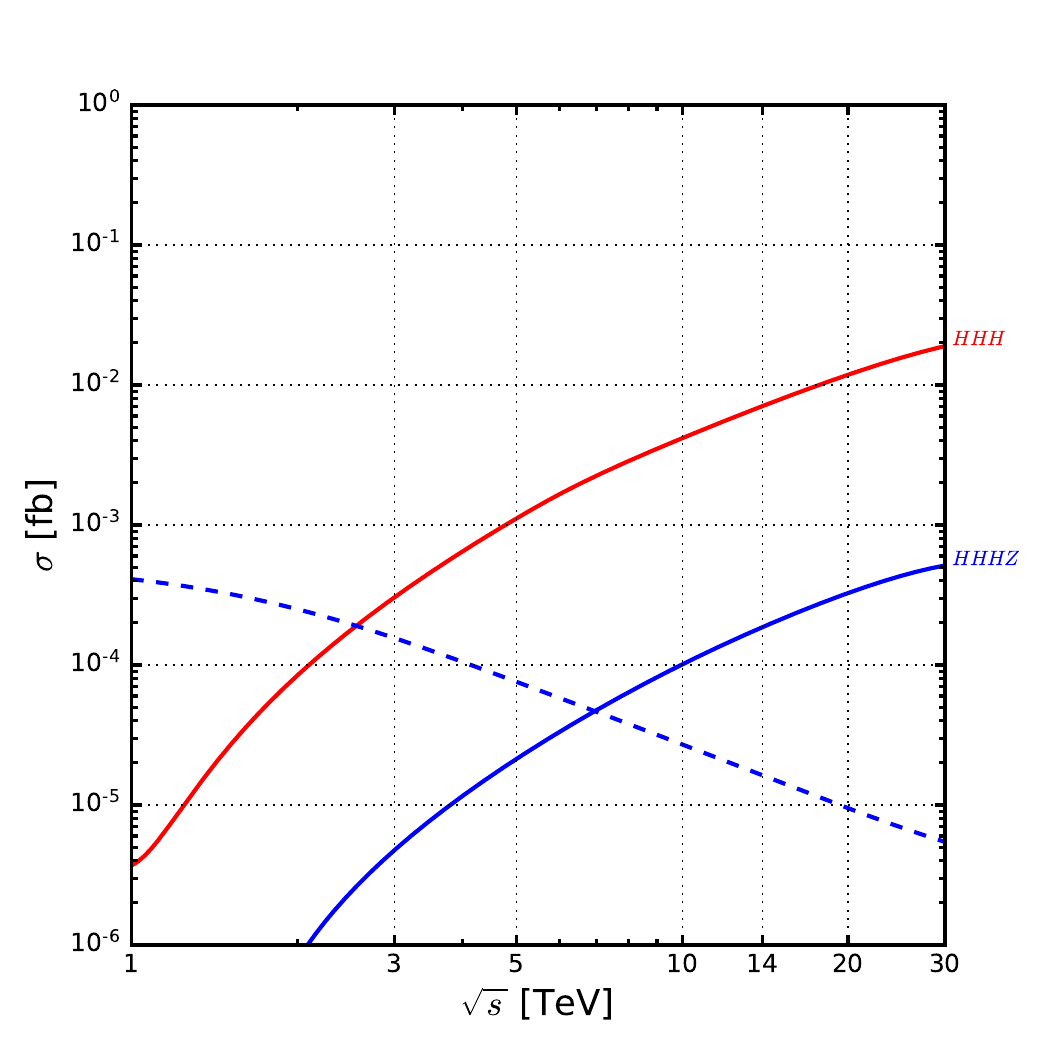}
    \caption{Energy dependence of the dominant single-Higgs (left), double-Higgs (centre) and triple-Higgs (right) production mechanisms at a muon collider, from Ref.~\cite{Costantini:2020stv}. Solid lines correspond to vector-boson fusion (VBF) processes, while dashed lines denote $s$-channel annihilation. The charged-current $WW$-fusion process becomes the dominant production mechanism above $\sqrt{s}\simeq3$~TeV, whereas Higgsstrahlung decreases approximately as $1/s$}
    \label{fig:h_hh_prod}
\end{figure*}

The energy dependence of the dominant Higgs production mechanisms is shown in Fig.~\ref{fig:h_hh_prod}, with single-Higgs production in the left panel and multi-Higgs production in the central and right panels. At low centre-of-mass energies, single-Higgs production is dominated by the Higgsstrahlung process, $\mu^+\mu^-\rightarrow ZH$. As the collision energy increases, however, vector-boson fusion (VBF) rapidly becomes dominant, since at multi-TeV energies the enhanced radiation of effectively light $W$ and $Z$ bosons makes the collider increasingly behave like a gauge-boson collider. In particular, the charged-current process $\mu^+\mu^-\rightarrow\nu_\mu\bar{\nu}_\mu H$ benefits from the increasing effective $W$-boson luminosity, leading to an approximately logarithmic increase of the production cross section, whereas the Higgsstrahlung cross section decreases approximately as $1/s$. As shown in Fig.~\ref{fig:h_hh_prod}, the transition occurs already below $\sqrt{s}\simeq3$~TeV.

A similar behaviour is observed for multi-Higgs production, as illustrated in the central and right panels of Fig.~\ref{fig:h_hh_prod}. Double- and triple-Higgs production are almost entirely driven by vector-boson fusion throughout the multi-TeV energy range, making a high-energy muon collider particularly well suited for direct studies of the Higgs self-interactions through measurements of the trilinear and quartic Higgs self-couplings.

\begin{table}[!b]
    \centering
    \caption{Standard Model Higgs production cross sections at a centre-of-mass energy of 10~TeV. The cross sections are obtained by interpolating the results of Ref.~\cite{Costantini:2020stv} according to the energy dependence presented in that work. The expected event yields are reported for an integrated luminosity of $20~\mathrm{ab}^{-1}$}
    \label{tab:higgs_muc_10tev}
    \begin{tabular}{lcc}
        \toprule
        Process & $\sigma$ [fb] & \makecell{Expected events\\for $20~\mathrm{ab}^{-1}$} \\
        \midrule
        $\mu^+\mu^- \rightarrow \nu_\mu\bar{\nu}_\mu H$
        & 840
        & $1.7\times10^{7}$ \\
    
        $\mu^+\mu^- \rightarrow \mu^+\mu^- H$
        & 68
        & $1.4\times10^{6}$ \\

        $\mu^+\mu^- \rightarrow \nu_\mu\bar{\nu}_\mu t\bar t H$
        & 0.12
        & $2.4\times10^{3}$ \\

        \midrule
        $\mu^+\mu^- \rightarrow \nu_\mu\bar{\nu}_\mu H\!H$
        & 3.6
        & $7.2\times10^{4}$ \\

        $\mu^+\mu^- \rightarrow \nu_\mu\bar{\nu}_\mu H\!H\!H$
        & $5.6\times10^{-3}$
        & 110 \\
        \bottomrule
    \end{tabular}
\end{table}

The baseline scenario of the International Muon Collider Collaboration (IMCC) envisages a muon collider operating at a centre-of-mass energy of $\sqrt{s}=10$~TeV, with two interaction points, each instrumented with a general-purpose detector~\cite{MuonColliderESPPU2025}. Each experiment is expected to collect an integrated luminosity of 10~ab$^{-1}$ over approximately five years of operation. The combined dataset of 20~ab$^{-1}$ would correspond to the production of approximately $2\times10^{7}$ Higgs bosons in single-Higgs processes and $7\times10^{4}$ Higgs boson pairs~\cite{Costantini:2020stv}.
These event samples enable precision measurements of Higgs couplings and rare decay modes, while the sizeable double\hyp{Higgs} sample provides direct sensitivity to the trilinear Higgs self-coupling. Although the production rate is considerably smaller, triple-Higgs production also becomes experimentally accessible, opening the possibility of probing the quartic Higgs self-coupling on a longer time scale. The benchmark Standard Model production cross sections corresponding to these processes are summarised in Table~\ref{tab:higgs_muc_10tev}.

The hierarchy of production cross sections naturally defines the experimental Higgs physics programme at a high-energy muon collider. The large single-Higgs sample is dominated by the charged-current vector-boson-fusion process, characterised by a Higgs boson recoiling against missing energy carried by the neutrinos. The complementary neutral-current fusion process, with two forward muons in the final state, provides a distinctive event topology and an independent probe of Higgs production. The same vector-boson-fusion topology also dominates double- and triple-Higgs production, providing a common experimental framework for studies of the Higgs self-interactions.

Given that Higgs boson production at multi-TeV energies is dominated by vector-boson fusion, its theoretical description must account for the electroweak radiation emitted by the incoming muons. This is naturally described within the framework of electroweak parton distribution functions (EW PDFs), which provide the probability density of finding a $W$ or $Z$ boson, or a photon, carrying a fraction of the muon momentum~\cite{Han:2022fup}. Unlike the proton PDFs used at hadron colliders, EW PDFs are perturbatively calculable and therefore introduce only a small, well-controlled theoretical uncertainty.

The dominant theoretical effects at multi-TeV energies arise from higher-order electroweak corrections. Owing to the large separation between the collision energy and the electroweak scale ($\sqrt{s}\gg m_W$), the virtual emission and reabsorption of electroweak gauge bosons generate logarithmically enhanced terms, known as electroweak Sudakov logarithms. These corrections increase in magnitude with the collision energy, modifying inclusive Higgs production cross sections by several percent, with effects reaching about $10\%$ in the high-energy tails of differential distributions. Consequently, precision Higgs measurements at a 10~TeV muon collider require theoretical predictions that include at least next-to-leading-order electroweak corrections together with the resummation of the leading logarithmically enhanced terms~\cite{Han:2022fup,AlAli:2022}.

\section{Analysis methodology}
\label{sec:methodology}

A common analysis framework is adopted for all Higgs production and decay channels considered in this review. The expected precision of the Higgs production cross-section measurements is evaluated using Monte Carlo (MC) simulated samples for both signal and background processes. Events are processed through a detailed simulation of the MUSIC detector, digitised, and reconstructed within the IMCC software framework. The event selection is optimised separately for each Higgs decay channel according to its experimental signature. Depending on the final state, either traditional cut-based selections or machine-learning techniques are employed to maximise the signal sensitivity and determine the expected measurement precision.

\subsection{Monte Carlo event generation, detector simulation, and reconstruction}
\label{sec:samples}

The Monte Carlo samples for the signal processes and the corresponding physics backgrounds were generated with \textsc{Whizard} v3.1.5~\cite{WHIZARD1,WHIZARD2}. Depending on the process, the \texttt{SM}, \texttt{SM\_ac}, and \texttt{SM\_HIGGS} models were employed. Parton showering and hadronisation were performed with \textsc{Pythia} v8.313~\cite{PYTHIA} using the configuration provided within \textsc{Whizard}. A loose event selection, less restrictive than the final analysis requirements, was applied at the generator level to improve the generation efficiency in the kinematic regions of interest. For each process, a sufficiently large event sample was generated so that the statistical uncertainty associated with the Monte Carlo samples is negligible.

For the sensitivity evaluation, the MC event samples are normalised according to
\[
w = \frac{\sigma_{\mathrm{prod}}\,\mathcal{B}\,\mathcal{L}_{\mathrm{int}}}{N_{\mathrm{gen}}}\ ,
\]
where $\sigma_{\mathrm{prod}}$ is the production cross section calculated with \textsc{Whizard}, $\mathcal{B}$ is the branching fraction into the selected final state taken from Ref.~\cite{LHCHiggsYR4}, $\mathcal{L}_{\mathrm{int}}$ is the assumed integrated luminosity, and $N_{\mathrm{gen}}$ is the total number of generated events. The use of a common event generator ensures a consistent theoretical description for all signal processes considered in this review.

The generated event samples are processed through the detailed MUSIC detector simulation based on\linebreak \textsc{Geant4} v11.1.0~\cite{GEANT}, and then reconstructed using the muon collider software framework~\cite{MuCSoftware}. The machine-induced background is overlaid onto the $\mu^+\mu^-$ hard-scattering events on an event-by-event basis prior to reconstruction. The two dominant sources of machine-induced background are considered~\cite{music-paper}: background from muon decay, conventionally referred to as beam-induced background (BIB), and background from incoherent $e^+e^-$ pair production (IPP). The BIB is simulated with the \textsc{Fluka} toolkit v4.4.0~\cite{FLUKA}, modelling the lattice and machine-detector interface of a 10~TeV collider in the ``EU24'' configuration~\cite{MDI-EU24}. The IPP background is generated with \textsc{Guinea-Pig} v1.4.3~\cite{GUINEA-PIG}, and the subsequent propagation of the resulting electrons and pos\-i\-trons into the detector is simulated with \textsc{Fluka}.

The MUSIC detector is a general-purpose detector concept optimised for operation at a multi-TeV muon collider in the presence of machine-induced background. It comprises an all-silicon tracking system, a high-gran\-u\-lar\-i\-ty crystal electromagnetic calorimeter, a sampling hadronic calorimeter, and an outer muon system. The tracking system and the electromagnetic calorimeter are immersed in a 5~T solenoidal magnetic field. Two conical tungsten shields surrounding the beam pipe\linebreak (``nozzles'') suppress the particle flux originating from beam-muon decays. A detailed description of the detector concept and its expected performance is provided in Ref.~\cite{music-paper}.

The reconstruction algorithms for physics objects have been optimised to mitigate the impact of machine-induced background while preserving the performance required for precision Higgs measurements. Charged-particle trajectories (tracks) are reconstructed with the ACTS tracking package~\cite{ACTS}, while energy deposits in the calorimeters are combined using the Pandora particle\hyp{flow} algorithm~\cite{PandoraPFA} to reconstruct individual particles. The resulting physics objects include photons, electrons, muons, and jets, which are used throughout the analyses presented in this work. The reconstruction algorithms and their expected performance are described in Ref.~\cite{music-paper}.
Photons are reconstructed from localised energy deposits in the electromagnetic calorimeter that are not associated with a charged-particle track, while electrons are identified by matching such energy deposits to reconstructed tracks. Since electron reconstruction relies on the combination of tracking and calorimeter information, it is intrinsically more challenging than photon or muon reconstruction, particularly in the presence of machine-induced background. Consequently, the current electron identification efficiency is lower than that expected after further algorithmic optimisation. Analyses relying heavily on electron final states should therefore be regarded as conservative, and some channels are deferred until improved electron reconstruction algorithms become available.\linebreak
Muons are identified by matching reconstructed tracks to signals in the outer muon system. Jets are reconstructed by clustering the particle-flow objects with the $k_t$ algorithm~\cite{kt} using a radius parameter of \mbox{$R = 0.5$}. Jets originating from $b$ quarks are identified with a dedicated flavour-tagging algorithm exploiting displaced secondary vertices and semileptonic heavy-flavour decays. Further details of the flavour-tagging procedure adopted for the $H\rightarrow b\bar{b}$ analysis are given in Sect.~\ref{subsec:Htobb}.
The reconstruction and identification of hadronically decaying $\tau$ leptons have not yet been implemented in the current reconstruction framework and are therefore not considered in this study.

Since the impact of machine-induced background on the reconstruction of physics objects has been extensively studied, Higgs boson decay channels with cleaner final states containing only photons or muons are simulated using a parametric detector simulation based on \textsc{Delphes} v3.5.0~\cite{DELPHES} in place of the full \textsc{Geant4}-based simulation, significantly reducing the required computing time.
The Muon Collider \textsc{Delphes} detector card has been tuned to reproduce the object-level performance of the detailed \textsc{Geant4}-based detector simulation, which includes the dominant machine-induced background.

\subsection{Signal extraction and sensitivity evaluation}

The sensitivity estimates for single-Higgs production, the Higgs boson mass, and double-Higgs production are presented in Sects.~\ref{sec:measurements}, \ref{sec:mass}, and \ref{sec:selfcoupling}, respectively. The determination of the trilinear Higgs boson self-coupling is discussed in Sect.~\ref{sec:trilinear}.
These studies share a common methodology and several underlying assumptions.

All studies assume a single experiment collecting an integrated luminosity of 10~ab$^{-1}$. Section~\ref{sec:xsec_summary} presents the expected statistical sensitivities obtained by extrapolating these results to the baseline two-experiment scenario described in Sect.~\ref{sec:physics10tev}.

The estimate of signal yields is based on the two dominant VBF production processes. 
Unless stated otherwise, the analyses are performed using only the WW-fusion sample, and the signal yields are rescaled by a factor of 1.103 to include the ZZ-fusion contribution, based on the ratio of the corresponding production cross sections calculated with \textsc{Whizard}. Dedicated simulated samples show that the WW- and ZZ-fusion processes have very similar kinematic distributions and selection efficiencies.

Although the analyses employ different kinematic selection criteria depending on the final state, all reconstructed physics objects are required to lie within the polar-angle range $10^\circ < \theta < 170^\circ$, ensuring good and uniform detector acceptance and reconstruction efficiency. 
Machine-learning techniques are employed,\linebreak whenever appropriate, to improve the separation of signal from the physics backgrounds and to identify jets originating from $b$ quarks.

The effective cross section for the considered decay channel is extracted as
\[
\sigma_{\mathrm{eff}}
\equiv \sigma_{\mathrm{prod}}\,\mathcal{B}
= \frac{N_S}{\epsilon_{\mathrm{sel}}\,\mathcal{L}_{\mathrm{int}}}\ ,
\]
where $N_S$ is the estimated number of signal candidates, $\epsilon_{\mathrm{sel}}$ is the signal selection efficiency, and $\mathcal{L}_{\mathrm{int}}$ is the integrated luminosity. 
In evaluating the expected sensitivity, the uncertainties on the selection efficiency and the integrated luminosity are assumed to be negligible, so that the expected measurement precision is determined entirely by the uncertainty on the extracted signal yield.

The signal yield is obtained from a fit to either the Higgs boson invariant-mass distribution or the output of a multivariate discriminant, depending on the decay channel. The statistical uncertainty on the signal yield is evaluated using toy Monte Carlo studies, in which pseudo-experiments are generated according to the expected signal and background models and analysed with the same fitting procedure used for the nominal analysis. The distribution of the fitted signal yields is then used to determine the expected statistical uncertainty.

\section{Sensitivity to Higgs production measurements}
\label{sec:measurements}

The expected sensitivity to the Higgs production cross sections is evaluated for the decay channels $H\to b\bar{b}$, $H\to WW^\ast\to q\bar{q}'\ell\nu_\ell$ ($\ell=e,\mu$), $H\to ZZ^\ast\to q\bar{q}\mu^+\mu^-$, $H\to\gamma\gamma$, and $H\to\mu^+\mu^-$, assuming the reference integrated luminosity of 10~ab$^{-1}$ discussed in Sect.~\ref{sec:methodology}. The results scaled to the combination of two experiments are summarised in Sect.~\ref{sec:xsec_summary}.

\subsection{$H\to b\bar{b}$}
\label{subsec:Htobb}

The $H \to b\bar{b}$ decay channel has the largest branching fraction of all Higgs boson decay modes and therefore plays a central role in the Higgs physics programme of a muon collider. 
Its identification and precise reconstruction are essential not only for measuring the Higgs coupling to $b$ quarks, but also for studies of double-Higgs production and the determination of the trilinear Higgs self-coupling. In particular, the $H\!H \to b\bar{b}b\bar{b}$ final state provides the highest event yield among Higgs pair decay channels. This section focuses on the measurement of the single-$H \to b\bar{b}$ process, while the study of the $H\!H \to b\bar{b}b\bar{b}$ channel is presented in Sect.~\ref{sec:selfcoupling}.

The background processes contributing to this Higgs decay mode comprise final states containing two jets not originating from an $H\to b\bar{b}$ decay, namely SM dijet production in association with two neutrinos ($qq\nu_\ell\bar{\nu}_\ell$), two charged leptons ($qq\ell\ell$) or one charged lepton and one neutrino ($qq\ell\nu_\ell$). In addition, the $H \to c\bar{c}$ decay constitutes an irreducible background to the $H \to b\bar{b}$ signal since the current analysis algorithm does not separate it from $H\to b\bar{b}$. Higgs decays into jets originating from light quarks ($u$, $d$, and $s$) or gluons are found to be negligible after application of the jet flavour-tagging procedure described below.

\begin{table}[t!]
    \centering
    \caption{Effective production cross sections, selection efficiencies, and expected event yields for signal and background processes in the $H \to b \bar{b}$ analysis with $\mu^+\mu^-$ collisions at 10 TeV and $\mathcal{L}_{\mathrm{int}}=10$ ab$^{-1}$}
    \label{tab:h2bb}
    \begin{tabular}{l|cccc}
    \toprule
    Process           &  $\sigma_{\text{eff}}$ [fb] & $\epsilon_{\text{sel}}$ [\%] & $N_{\text{exp}}$ \\
    \midrule
    $H \to b \bar{b}$ &    532         &     10.9                     & 581120 \\
    \midrule
      $H \to c \bar{c}$ &     26.4       &      1.36                    &   3606 \\ 
    $qq\nu_\ell\bar{\nu}_\ell$  &   1052         &      5.24                    & 551559 \\
    $qq\ell\nu_\ell$  &   9763         &      0.0342                  &  33388 \\     
    $qq\ell\ell$      &   4339         &      0.0491                  &  21308 \\   
    \bottomrule
    \end{tabular}
\end{table}

Candidate $H \to b\bar{b}$ events are selected by requiring two reconstructed jets of radius parameter $R=0.5$ with transverse momentum $p_T^j > 40$~GeV, reducing the low-energy component of the combinatorial background. Jets originating from $b$ quarks are identified using the flavour-tagging algorithm described in Ref.~\cite{music-paper}.
A jet is labelled as \emph{tagged} if it satisfies one of the following requirements:
\begin{itemize}
    \item a reconstructed secondary vertex (SV) together with a muon with $p_T^\mu>2$~GeV;
    \item  only a muon consistent with a semileptonic heavy-flavour decay;
    \item an SV without an associated muon.
\end{itemize}
In the latter case, the jets are subsequently classified using two deep neural networks (DNNs): one optimised to discriminate $b$ jets from light-flavour jets and the other to separate $b$ jets from $c$ jets. Both networks exploit input variables describing the properties of the jet and the reconstructed secondary vertex. 
The most discriminating observables are:
\begin{itemize}
\item the transverse and three-dimensional displacements of the SV from the primary interaction vertex;
\item the significance of the SV fit;
\item the SV proper decay time;
\item the invariant mass of the tracks associated with the SV;
\item the number of tracks associated with the SV and their displacements;
\item the reconstructed transverse momentum of the SV;
\item the fraction of the jet transverse momentum carried by the SV.
\end{itemize}
Selection requirements are applied to the outputs of both DNNs. The working point adopted for the $H \to b\bar{b}$ analysis is optimised to maximise the $b$-jet selection efficiency while minimising the misidentification probability for $c$ jets and light-flavour jets. 
The resulting overall $b$-tagging efficiency, including all the tagging categories, is about $60\%$, with misidentification probabilities of $22\%$ for $c$ jets and $0.9\%$ for light-flavour jets. Both jets forming the Higgs boson candidate are required to be $b$-tagged.

The signal and background samples are generated, simulated, and reconstructed following the procedure described in Sect.~\ref{sec:samples}. The corresponding selection efficiencies are determined from these samples, and the expected event yields for an integrated luminosity of 10 ab$^{-1}$ are obtained by normalising to the effective cross sections. Table~\ref{tab:h2bb} summarises the cross sections, selection efficiencies, and expected event yields. 
The dominant background is the $qq\nu_\ell\bar{\nu}_\ell$ process, whose expected event yield is comparable to that of the $H \to b \bar{b}$ signal. This background is dominated by $Z$ boson production with subsequent hadronic decays. Therefore, the signal extraction relies on the dijet invariant mass, which provides the main discrimination between signal and background.
 
The dijet invariant mass distributions for both signal and background processes are parameterised by fitting the corresponding simulated samples with double-Gaussian probability density functions.  The resulting probability density functions are normalised to the expected event yields reported in Table~\ref{tab:h2bb} and combined to construct the likelihood model.
Pseudo-experiments are then generated from this model, and each pseudo-dataset is fitted using the same parameterisation to extract the signal yield.

\begin{figure}[t!]
    \centering
    \includegraphics[width=\columnwidth]{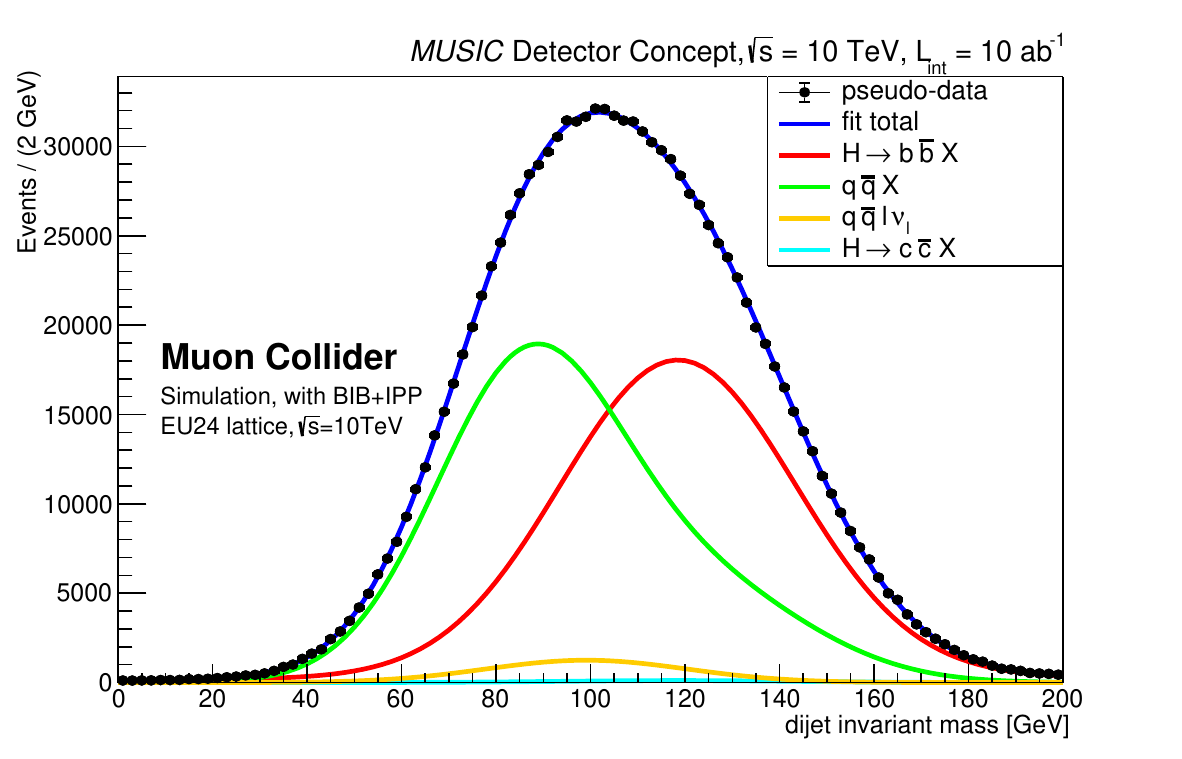}
    \caption{Example of a fit to the dijet invariant-mass distribution used to extract the $H\to b\bar b$ signal yield. The pseudo-data are shown as black points, while the solid blue line represents the total fit. The individual contributions from the $H\to b\bar b$ signal and the $q\bar qX$, $q\bar q\ell\nu_\ell$, and $H\to c\bar c$ backgrounds are shown separately}
    \label{fig:h2bb_fit}
\end{figure}

In the fit, the signal yield ($H \to b \bar{b}$) and the yields of the $qqX$ background components, with $X=\nu_\ell\bar{\nu}_\ell$ or $\ell\ell$, are treated as free parameters, 
while the contributions from the $qq\ell\nu_\ell$ and $H\to c\bar c$ processes are fixed to their expected values. A total of $1\,000$ pseudo-experiments are generated and fitted, and the average statistical uncertainty on the fitted signal yield is extracted. An example of a fit is shown in Fig.~\ref{fig:h2bb_fit}.

This uncertainty is interpreted as the statistical uncertainty on the $H \to b \bar{b}$ cross section measurement, resulting in a relative statistical uncertainty of
\begin{equation}
    \frac{\Delta \sigma (H \to b\bar{b})}{\sigma (H \to b\bar{b})} = 0.25\%\ .
\end{equation}
Given the importance of the $H \to b\bar{b}$ channel and the excellent statistical precision expected for its cross\hyp{section} measurement, the impact of the associated systematic uncertainties needs to be discussed. The dominant systematic effects are expected to arise from the determination of the jet energy scale (JES) and jet energy resolution (JER).
The jet energy calibration, \emph{i.e.} the procedure used to map the raw jet momentum, obtained by summing the reconstructed momenta of the jet constituents, to the true jet momentum, is initially derived from simulation, where the generator-level jet four-momentum is available. However, once real data become available, differences between data and simulation may lead to residual mismodelling, making data-driven calibration samples necessary. These residual corrections are commonly factorised into jet energy scale and jet energy resolution corrections. An estimate of the systematic uncertainties related to these corrections is given in the following, using the available simulation samples.
For $b$ jets, the same $b\bar{b}$ sample used for the extraction of the $H \to b\bar{b}$ signal can also be exploited to determine these corrections. Indeed, both the $Z$- and Higgs-boson resonances are present in this sample, allowing the calibration to be constrained over a broad range of jet transverse momenta. In the fit described in this section, the JES and JER corrections are introduced as nuisance parameters and are simultaneously determined together with the signal yield. Their impact on the $H \to b\bar{b}$ cross-section measurement can therefore be evaluated directly. The increase in the signal-yield uncertainty relative to the nominal fit, assuming the various contributions can be combined in quadrature, is interpreted as the corresponding systematic uncertainty.

The fit determines the jet energy scale and jet energy resolution corrections with relative uncertainties of 0.06\% and 0.10\%, respectively. The jet energy scale uncertainty translates into a 0.49\% uncertainty on the extracted $H\to b\bar{b}$ yield, while the contribution from the jet energy resolution is negligible compared with the statistical uncertainty.
With the current analysis strategy, the overall uncertainty on the $H \to b\bar{b}$ measurement is therefore expected to be dominated by the jet energy scale uncertainty. Nevertheless, once collider data become available, dedicated calibration channels will be exploited to constrain this systematic uncertainty, thereby further improving the overall measurement precision.

\subsection{$H\to WW^{\ast}$}

The $H\to WW^\ast$ channel is studied in the semileptonic final state, where one $W$ boson decays into a muon or an electron and the corresponding neutrino, while the other decays hadronically into two jets, yielding the final states $q\bar{q}^\prime\mu \nu_\mu$ and $q\bar{q}^\prime e \nu_e$.

\begin{table}[!t]
    \centering
    \caption{Effective production cross sections, selection efficiencies, and expected event yields for signal and background processes in the $H \to WW^{*}$ analysis  at a $\sqrt{s}=10$~TeV muon collider  assuming an integrated luminosity of $\mathcal{L}_{\mathrm{int}}=10$~ab$^{-1}$}
    \label{tab:h2ww}
    \begin{tabular}{l c | c c c}
    \toprule
    Process & Channel & $\sigma_{\text{eff}}$ [fb] & $\epsilon_{\text{sel}}$ [\%] & $N_{\text{exp}}$ \\
    \midrule
    \multirow{2}{*}{$H \to WW^{*}$} & $\mu$ & \multirow{2}{*}{29.0}   & 49.0 &  142421 \\
                                    & $e$   &                      & 42.5 & $123470$ \\
    \midrule
    \multirow{2}{*}{$H\,\nu_\ell\bar\nu_\ell$} & $\mu$ & \multirow{2}{*}{846}   & ${<}~0.012$ & ${<}~984$ \\
                                               & $e$   &                        & 19.8 & $1624145$ \\
    \addlinespace
    \multirow{2}{*}{$H\,\ell\ell$} & $\mu$ & \multirow{2}{*}{87.5}   & 7.9  & $67066$ \\
                                   & $e$   &                       & 21.0 & $177833$ \\
    \addlinespace
    \multirow{2}{*}{$q\bar{q}^\prime\,\ell\nu_\ell$} & $\mu$ & \multirow{2}{*}{9772} & 2.7  & 2652923 \\
                                              & $e$   &                       & 17.1 & 16739401 \\
    \addlinespace
    \multirow{2}{*}{$q\bar{q}^\prime\,\nu_\ell\bar\nu_\ell$} & $\mu$ & \multirow{2}{*}{2674} & 6.9  & 1831630 \\
                                                      & $e$   &                       & 20.2 & 5408146 \\
    \addlinespace
    \multirow{2}{*}{$q\bar{q}^\prime\,\ell\ell$} & $\mu$ & \multirow{2}{*}{4339} & 3.1 & 1371037 \\
                                          & $e$   &                       & 2.1 & 891494 \\
    \bottomrule
    \end{tabular}
\end{table}

The analyses of the muon and electron channels are performed in two stages. First, a loose preselection is applied to select candidate Higgs boson events and to suppress a fraction of the low-energy backgrounds. Then, Boosted Decision Trees (BDTs)~\cite{BDTpython} are employed to improve the separation between signal and background processes.
Candidate $H\to WW^*$ events are reconstructed by combining a high-$p_T$ lepton with two reconstructed jets. 
Leptons are required to have $p_T^\ell > 10$ GeV.
If more than one identified lepton satisfies these requirements, the candidate with the largest separation from the jets associated with the hadronic $W$ boson,  defined by  $\Delta R = \sqrt{\Delta\phi^2 + \Delta\eta^2}$, is selected.
In the electron channel, events containing a muon satisfying the same selection criteria are vetoed.
Jets are required to satisfy $p_T^{j} > 20$ GeV. 
If more than two jets pass this selection, the jet pair with an invariant mass closest to the nominal $W$ boson mass is chosen as the hadronic $W$ boson candidate.
The expected number of events for the $H \to WW^*$ signal and the dominant background contributions after the preselection is summarised in Table~\ref{tab:h2ww}.

\begin{figure*}[!t]
    \centering
    \includegraphics[width=1\linewidth]{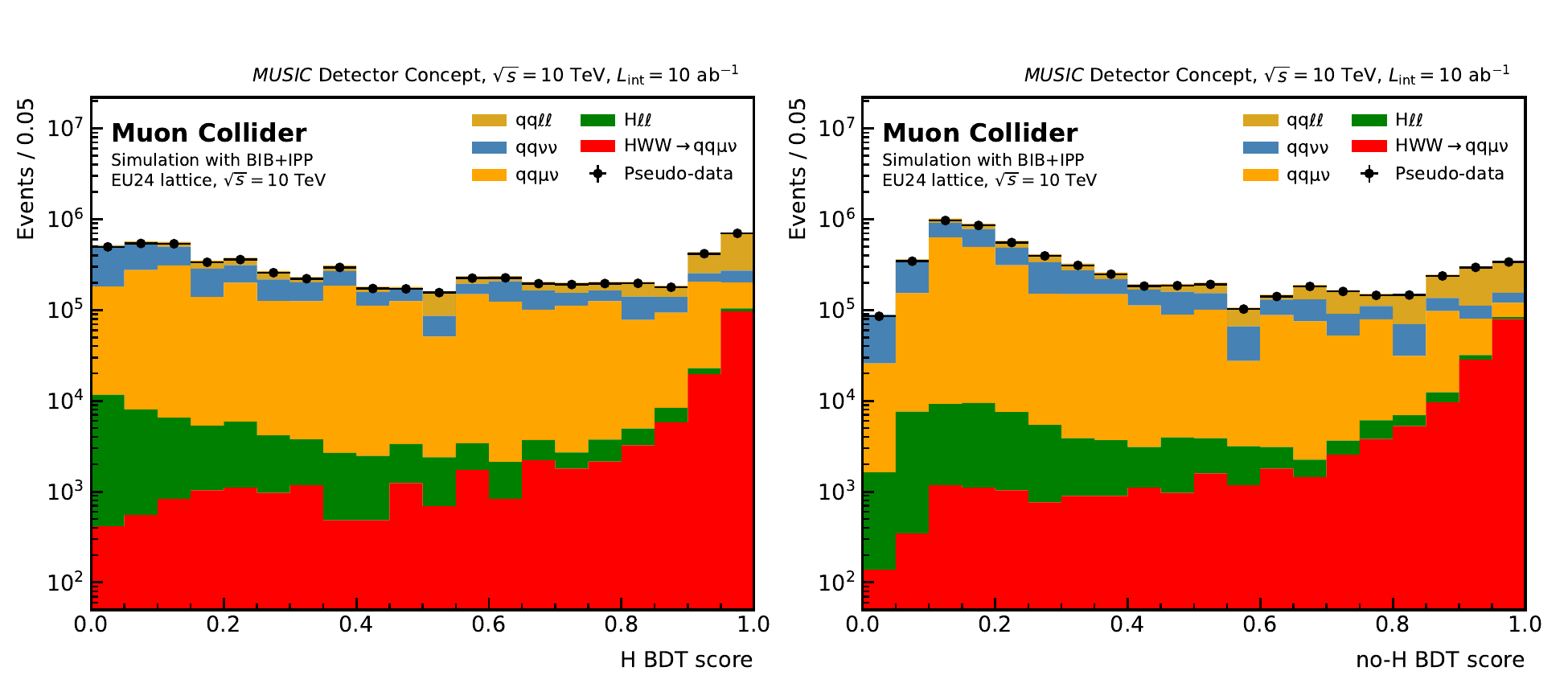} \\
    \includegraphics[width=1\linewidth]{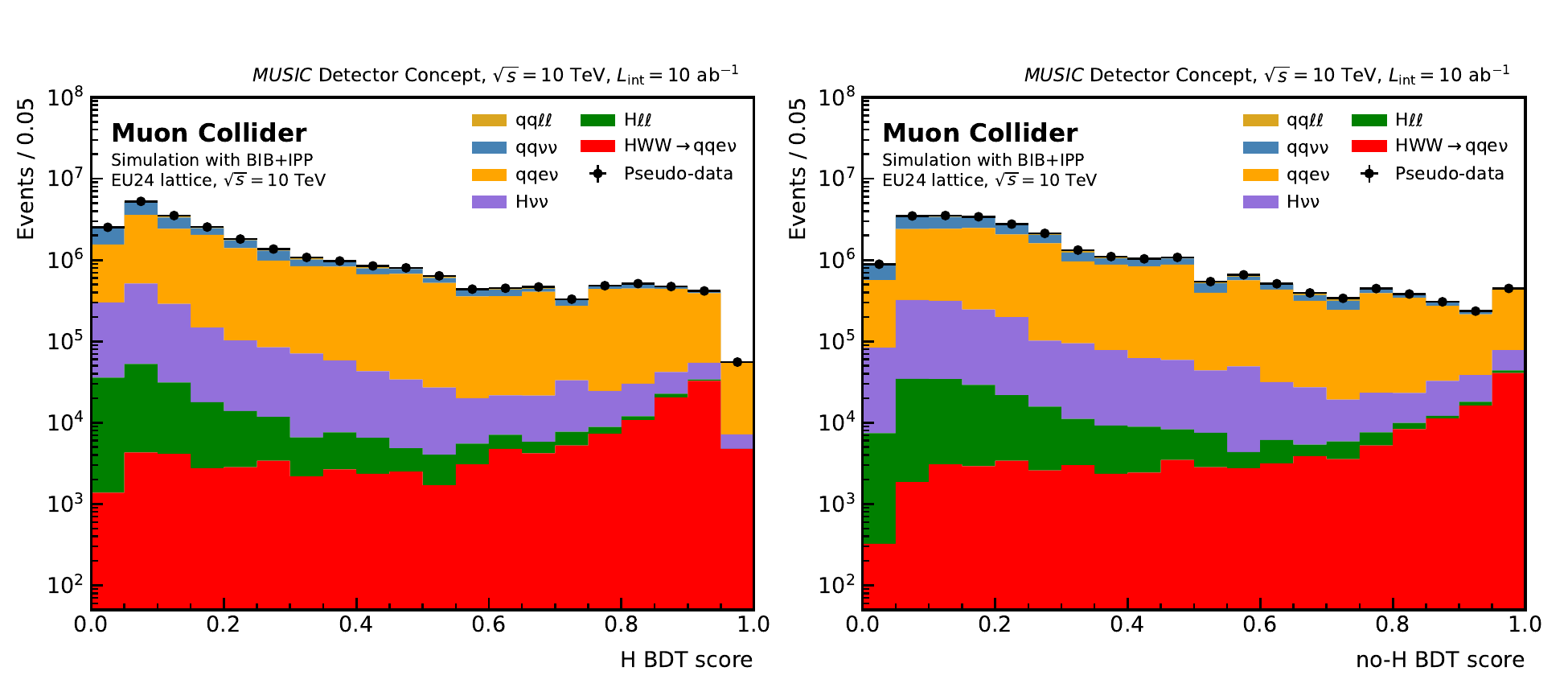}
    \caption{Projections of the two-dimensional fit for one pseudo-experiment onto the $H\!X$ (left) and $q\bar qX$ (right) BDT output axes. The upper and lower panels correspond to the muon and electron channels, respectively. The pseudo-data are shown as points and are overlaid on the stacked signal and background templates used in the fit
    \label{fig:h2ww}}
\end{figure*}

After the initial preselection, two BDT classifiers are trained independently for each leptonic channel to distinguish the signal from the Higgs boson backgrounds ($H\!X$) and the non-resonant backgrounds ($q\bar{q}X$), respectively. The BDTs exploit kinematic and topological properties of the reconstructed physics objects, including:
\begin{itemize}
    \item \textbf{lepton:} transverse momentum, polar angle, transverse impact parameter, the number of isolated leptons, the separations $\Delta R$ between the lepton and each of the two jets, and the minimum of the two $\Delta R$ values;
    \item \textbf{jets:} transverse momentum and polar angle of the two jets, and total number of jet constituents;
    \item \textbf{hadronic $W$ candidate:} transverse momentum, polar angle, and invariant mass, reconstructed from the two selected jets;
    \item \textbf{Higgs boson candidate:} polar angle, invariant mass, and the acollinearity angles between the lepton and the $W$ boson, the lepton and the Higgs boson, and the $W$ boson and the Higgs boson;
    \item \textbf{global event observables:} missing transverse momentum and the scalar sum of the transverse momenta of the lepton and the two jets.
\end{itemize}
The number of isolated leptons is defined as the number of leptons in the event with a minimum separation of $\Delta R>0.5 $ from both selected jets.
The acollinearity between two objects is computed from their momenta as
\[
    A_{1,2} = \pi - \arccos \frac{\vec{p}_1 \cdot \vec{p}_2}{|\vec{p}_1|\,|\vec{p}_2|}\ ,
\]
while the missing transverse momentum is given by
\[
    p_{T}^{\text{miss}} = \sqrt{p_x^2+p_y^2} \ ,
\]
where $p_{x,y} = - p_{x,y}^\ell - \sum_{j} p_{x,y}^{j}$.

The statistical precision of the signal yield in the two channels is evaluated using a toy Monte Carlo study, which fits the two-dimensional distribution of the $H\!X$ and $q\bar{q}X$ BDT outputs.
Two extended binned maximum\hyp{likelihood} models are defined with three components: the signal $H\to WW^\ast$, backgrounds involving a Higgs boson, and non-resonant backgrounds. The $2D$ histograms serve as probability density functions for the likelihood.
The bin widths are chosen to ensure approximately ten events per bin, thereby avoiding sparsely populated bins in the tails of the distributions.
A total of 10\,000 pseudo-experiments are generated from the likelihood model and fitted to extract the signal yield. The expected statistical precision is determined from the width of the fitted signal-yield distribution.
Figure~\ref{fig:h2ww} shows, for one pseudo-experiment, the projections of the two-dimensional fit onto the \(H\!X\)-BDT and \(q\bar qX\)-BDT output axes for the muon and electron channels.
Despite the unfavourable signal-to-background ratio after the preselection, the two BDT classifiers provide sufficient discrimination in the $H\!X$--$q\bar{q}X$ score plane to achieve an expected statistical precision of 0.55\% and 0.58\% on the signal yield in the muon and electron channels, respectively.

The combination of the muon and electron channels is performed within the same extended binned maximum\hyp{likelihood} framework. Instead of determining the signal yields, a common signal-strength parameter is introduced, defined as the ratio of the measured signal yield to the SM expectation,
$\mu = N_{\text{meas}} / N_{\text{SM}}$, such that $\mu = 1$ corresponds to the SM prediction.
A simultaneous fit to the two-dimensional $H\!X$ and $q\bar{q}X$ BDT distributions in the two channels is then performed, with $\mu$ treated as a common parameter of the two likelihoods, while the background normalisations are allowed to vary independently. 
The expected relative statistical uncertainty on the signal strength, equivalent to that on the $H\to WW^\ast$ production cross section, is
\begin{equation}
    \frac{\Delta \sigma (H \to WW^\ast)}{\sigma (H \to WW^\ast)}= 0.49\%\ .
\end{equation}

\subsection{$H\to ZZ^{\ast}$}

The $H\to ZZ^\ast$ decay is reconstructed in the semileptonic final state, where one $Z$ boson decays into two isolated, oppositely charged muons and the other decays hadronically into two jets, yielding the final state $\mu^+\mu^-q\bar{q}$.

\begin{table}[!t]
    \centering
    \caption{Effective production cross sections, selection efficiencies, and expected event yields for signal and background processes in the $H \to ZZ^\ast$ analysis with $\mu^+\mu^-$ collisions at 10 TeV and $\mathcal{L}_{\mathrm{int}}=10$ ab$^{-1}$}
    \label{tab:h2zz}
    \begin{tabular}{l|cccc}
    \toprule
    Process   & $\sigma_{\text{eff}}$ [fb] & $\epsilon_{\text{sel}}$ [\%] & $N_{\text{exp}}$ \\
    \midrule
    $H\rightarrow ZZ^\ast\rightarrow\mu^+\mu^- q\bar{q}$  &  1.04  &  6.2   &  643 \\
    \midrule
    $\mu^+\mu^- qq$                                       &  41.6  &  5.3   &   22107 \\
    $\mu^+\mu^- \ell^+\ell^-$                                   &  0.102  &  1.3   &   13 \\
    \bottomrule
    \end{tabular}
\end{table}

\begin{figure*}[t]
    \centering
    \includegraphics[width=0.49\linewidth]{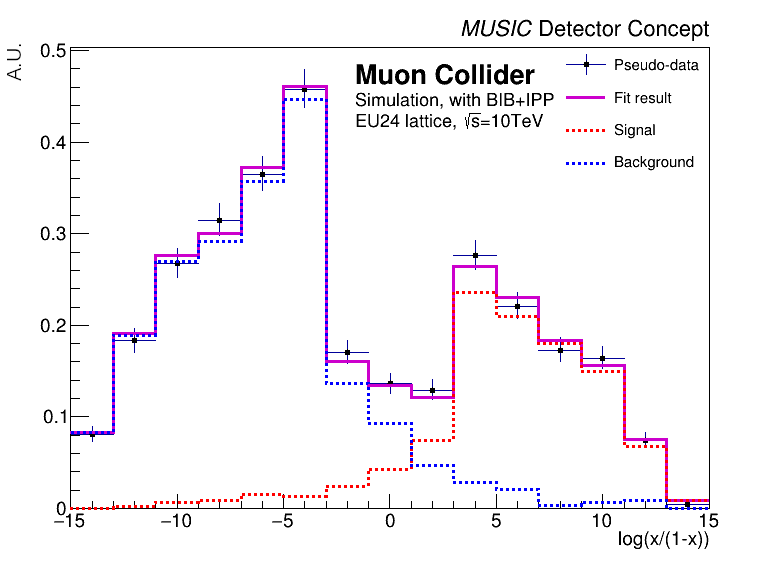}
    \hfill
    \includegraphics[width=0.49\linewidth]{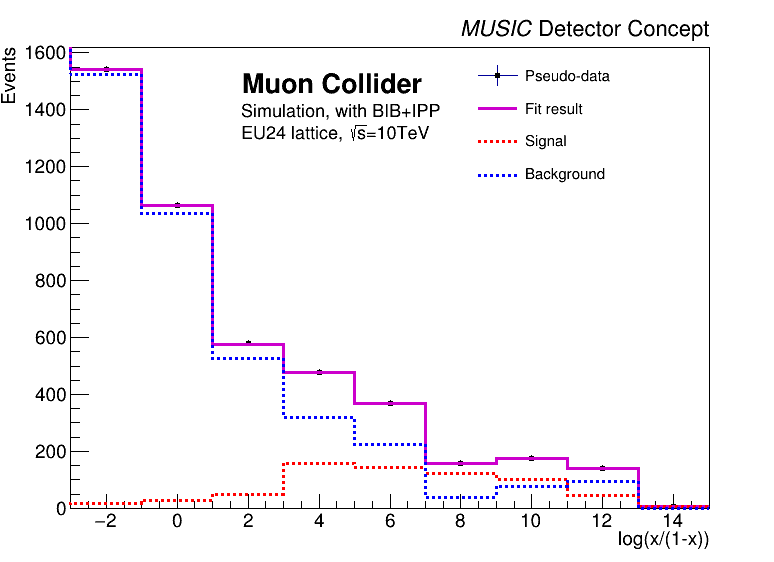}
    \caption{BDT distributions for the $H\to ZZ^\ast$ analysis. The total fit is shown by the solid purple line, while the signal and background components are represented by the red and blue dotted lines, respectively. The left panel shows the distributions nor\-mal\-ised to unit area, while the right panel shows the distributions normalised to the expected event yields
    \label{fig:h2zz}}
\end{figure*}

The analysis is performed in two stages. A loose preselection is first applied to identify candidate $H\to ZZ^\ast$ events and suppress the low-energy background. Subsequently, a BDT classifier is employed to improve the separation between signal and background processes.
Events are selected by requiring at least two reconstructed muons with opposite electric charge and transverse momentum $p_T^\mu > 20$~GeV, together with at least two reconstructed jets satisfying $p_T^j > 10$~GeV.
The two highest-$p_T$ muons are combined to reconstruct the leptonic $Z^{(*)}$ boson candidate, while the two highest-$p_T$ jets are combined to reconstruct the hadronic $Z^{(*)}$ boson candidate.
The invariant masses of the leptonic and hadronic candidates are required to be consistent with the expected $Z$ boson mass within an optimised selection window, suppressing combinatorial and non-resonant backgrounds.
The dominant irreducible background arises from continuum $Z\!Z^*$ production with subsequent decays to $\mu^+\mu^-q\bar{q}$. Additional background contributions originate from processes such as $Z+$jets, $W\!Z$, and multiboson production, as well as other electroweak processes that can mimic the signal topology through misreconstruction or additional jet activity. 
The expected signal and background event yields after the preselection are summarised in Table~\ref{tab:h2zz}.

To enhance the separation between signal and background processes, a multivariate analysis based on a Gradient Boosted Decision Tree (BDT-G)~\cite{TMVA} is employed. The classifier is trained using a set of kinematic observables sensitive to the signal topology, including:
\begin{itemize}
    \item the invariant masses of the muon and jet pairs; 
    \item the transverse momenta of the reconstructed leptonic and hadronic $Z$ candidates;
    \item the angular variables $\cos\theta_{ZZ}$, $\cos\theta_{\mu Z}$, $\cos\theta_{j Z}$, and $\cos\theta_{ZH}$, describing the relative orientations of the reconstructed final-state candidates;
    \item the reconstructed Higgs boson mass;
    \item global event observables: the missing transverse momentum and the scalar sum of the transverse momenta of the muons and jets.
\end{itemize}
The output of the trained BDT-G classifier is transformed according to
\[
    D=\log\frac{x}{1-x}\ ,
\]
where $x=(BDT+1)/2$.
This transformation expands the signal-enriched region and improves the discrimination power between signal and background templates. Normalised signal and background templates are constructed from the transformed discriminant distribution and subsequently used in the statistical analysis. Figure~\ref{fig:h2zz} shows the BDT output distributions normalised to unit area (left) and to the expected event yields (right).

The expected sensitivity is determined by using a binned maximum-likelihood fit to the transformed BDT discriminant distribution. The likelihood model consists of signal and background template shapes, with independent normalisation parameters for the signal strength and background yield. To estimate the expected statistical precision, ensembles of pseudo-ex\-per\-i\-ments are generated. For each pseudo-experiment, the expected signal and background yields are fluctuated according to Poisson statistics on a bin-by-bin basis to create pseudo-data samples. The maximum-likelihood fit is then performed, yielding the best-fit value of the signal-strength parameter and its associated uncertainty.

A total of 10\,000 pseudo-experiments are generated, and the relative signal-strength uncertainty is evaluated for each pseudo-experiment. The expected statistical precision is taken from the median of the resulting uncertainty distribution. The analysis yields an expected relative statistical uncertainty on the measurement of the $H\rightarrow ZZ^*\rightarrow\mu^+\mu^- q\bar{q}$ production cross section of
\begin{equation}
    \frac{\Delta \sigma (H \to ZZ^\ast)}{\sigma (H \to ZZ^\ast)} = 6.0\%\ .
\end{equation}

\subsection{$H\to \gamma\gamma$}
\label{subsec:Htogammagamma}

The $H\to\gamma\gamma$ decay is characterised by the presence of two isolated high-energy photons with a narrow invariant\hyp{mass} peak centred at the Higgs boson mass, providing one of the cleanest experimental signatures despite its small branching fraction.
The signal and background samples are processed using the \textsc{Delphes} parametric simulation of the MUSIC detector.

The analysis begins with a loose preselection to identify candidate $H\to\gamma\gamma$ events and suppress the low-energy background.
Events are required to contain at least two reconstructed photons with transverse momentum $p_T^\gamma > 20\ {\rm GeV}$.
The two highest-$p_T$ photons are combined to form the Higgs boson candidate. A loose requirement on the diphoton invariant mass, $50 < m_{\gamma\gamma} < 150\ {\rm GeV}$, is applied to retain the signal region while suppressing a large fraction of the non-resonant background.
\begin{table}[b!]
    \centering
    \caption{Effective cross sections, selection efficiencies, and expected event yields for the signal and background processes in the $H\to\gamma\gamma$ analysis for $\mu^+\mu^-$ collisions at $\sqrt{s}=10$~TeV, assuming an integrated luminosity of $\mathcal{L}_{\mathrm{int}}=10$~ab$^{-1}$}
    \label{tab:h2aa}
    \begin{tabular}{l|cccc}
    \toprule
    Process   & $\sigma_{\text{eff}}$ [fb] & $\epsilon_{\text{sel}}$ [\%] & $N_{\text{exp}}$ \\
    \midrule
    $H\nu_\ell\bar{\nu}_\ell,\: H \to \gamma\gamma$ &  1.95      &  44.0     &  8555 \\
    $H\ell\ell,\: H \to \gamma\gamma$               &  0.211     &  44.6     &  939 \\
    \midrule
    $\nu_\ell\bar{\nu}_\ell\gamma\gamma$            &  9126   &  0.18   &  164898 \\
    $\ell\ell\gamma\gamma$                          &  0.806      &  6.1     &  490 \\   
    $\ell\ell\gamma$                                &  21.9     &  0.03     &  66 \\
    \bottomrule
    \end{tabular}
\end{table}
The expected signal and background event yields after the preselection are summarised in Table~\ref{tab:h2aa}.
The dominant irreducible backgrounds originate from the continuum processes $\mu^+\mu^- \rightarrow \gamma\gamma\nu_\ell\bar{\nu}_\ell$ and $\mu^+\mu^- \rightarrow \gamma\gamma\ell^+\ell^-$. Additional reducible contributions arise from processes involving jets misidentified as photons. 
A sample of $\mu^+\mu^-\to\gamma\gamma$ events was also simulated, but no events satisfied the preselection requirements, making this background negligible.

The signal yield is extracted from an unbinned extended maximum-likelihood fit to the diphoton invariant\hyp{mass} distribution. The signal component is parameterised by a double Crystal Ball function, which accurately describes the detector resolution and non\hyp{Gaussian} tails, while the background contribution is modelled using a polynomial function. 

\begin{figure}[t!]
    \centering
    \includegraphics[width=\columnwidth]{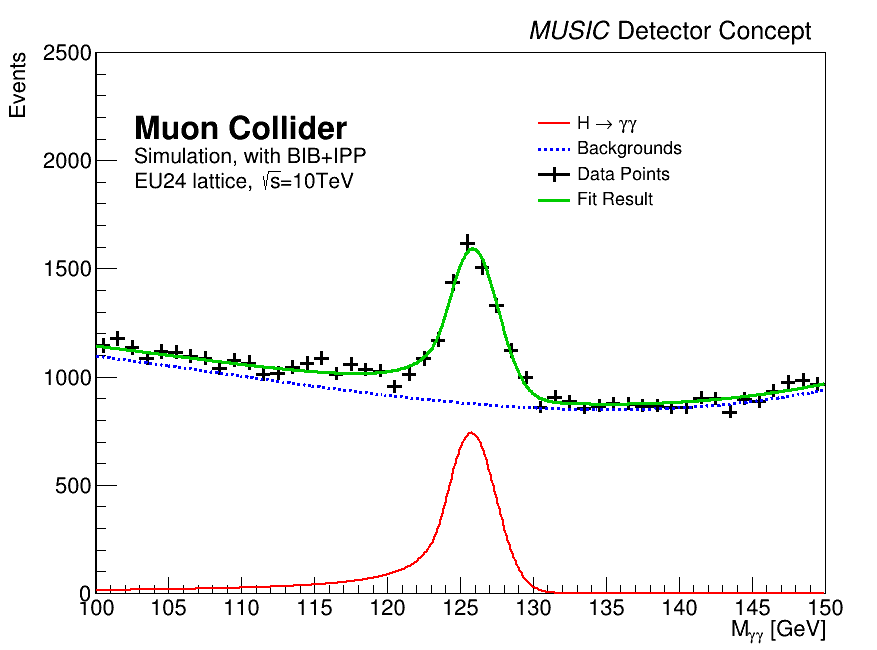}
    \caption{Diphoton invariant-mass distribution for a representative pseudo-experiment (points), with the total fit (blue solid line) superimposed and the $H\to\gamma\gamma$ contribution (red solid line)
    \label{fig:h2aa}}
\end{figure}

To estimate the expected statistical precision, ensembles of pseudo-experiments are generated. For each of them, the expected signal and background yields are fluctuated according to Poisson statistics to construct pseudo-data samples. Each pseudo-dataset is fitted with the same unbinned extended maximum-like\-li\-hood model, extracting the best-fit value of the signal strength parameter and its associated uncertainty.
A total of 10\,000 pseudo-experiments are generated, and the relative signal strength uncertainty is evaluated from each of them. The expected statistical precision is taken as the median of the corresponding distribution of relative uncertainties. 
Figure~\ref{fig:h2aa} shows the diphoton invariant-mass distribution for a representative pseudo-experiment, together with the total fit and the fitted $H\to\gamma\gamma$ and background components. The $H\to\gamma\gamma$ signal is clearly visible as a narrow peak near the Higgs boson mass.

The analysis yields an expected relative statistical uncertainty on the measurement of the $H\rightarrow\gamma\gamma$ production cross section of
\begin{equation}
    \frac{\Delta \sigma (H \to \gamma\gamma)}{\sigma (H \to \gamma\gamma)} = 3.7\%\ .
\end{equation}

\subsection{$H\to \mu^+\mu^-$}
\label{subsec:Htomumu}

The $H\to\mu^+\mu^-$ decay provides direct access to the Higgs boson coupling to second-generation fermions. Despite its small branching fraction, of the order of $10^{-4}$, the clean dimuon signature and the large Higgs boson production rate at a 10~TeV muon collider enable a precise measurement of its production cross section.

The $H \to \mu^{+}\mu^{-}$ analysis is performed using the \textsc{Delphes} parametric simulation of the MUSIC detector. 
Events are required to contain at least two oppositely charged muons with $p_{T}^{\mu} > 10$~GeV. The contribution from fake muons induced by machine-induced background is modelled by injecting additional muons that satisfy the same kinematic requirements, sampled from their expected transverse-momentum and polar-angle distributions; the injection rate of $0.8\%$ per event is taken from the detailed \textsc{Geant4} detector simulation. 
If more than two oppositely charged muons are present, the $\mu^{+}\mu^{-}$ pair with invariant mass closest to the nominal Higgs boson mass is retained. 
A loose preselection is then applied by requiring the scalar sum of the muon transverse momenta, the dimuon invariant mass, and the dimuon transverse momentum to satisfy $\sum p_{T}^{\mu} > 100$~GeV, $105 < m_{\mu\mu} < 145$~GeV, and $p_{T}^{\mu\mu} > 30$~GeV, respectively. After this preselection, 639 signal events are expected, together with the two dominant background processes, $\mu^{+}\mu^{-} \to \mu^{+}\mu^{-}\mu^{+}\mu^{-}$ (43\,128 events) and $\mu^{+}\mu^{-} \to \mu^{+}\mu^{-}\nu_{\mu}\bar{\nu}_{\mu}$ (17\,108 events).

\begin{table}[b!]
    \centering
    \caption{Effective production cross sections, selection and BDT-cut efficiencies, and expected event yields for signal and background processes in the $H \to \mu^+\mu^-$ analysis with $\mu^+\mu^-$ collisions at 10 TeV and $\mathcal{L}_{\mathrm{int}}=10$ ab$^{-1}$}
    \label{tab:h2mumu}
    \begin{tabular}{l|cccc}
    \toprule
    Process                  &  $\sigma_{\text{eff}}$ [fb] & $\epsilon_{\text{sel}}$ [\%] & $\epsilon_{\text{BDT}}$ [\%] & $N_{\text{exp}}$ \\
    \midrule
    $H \to \mu^+\mu^-$ &  0.202                   &      31.6                    &  80.6                        & 515   \\
    \midrule
    $\mu^+\mu^-\mu^+\mu^-$ & 156                  &      2.8                     &  48.6                        & 20956 \\
    $\mu^+\mu^-\nu_\mu\bar{\nu}_\mu$ & 1536       &      0.11                    &  71.1                        & 12160 \\   
    \bottomrule
    \end{tabular}
\end{table}

Two boosted decision trees~\cite{TMVA}, $\mathrm{BDT}(4\mu)$ and\linebreak $\mathrm{BDT}(2\mu2\nu)$, are trained on independent samples to separate the signal from each background component. 
The classifiers exploit the following observables:
\begin{itemize}
    \item \textbf{muon kinematics:} transverse momentum of the subleading muon;
    \item \textbf{dimuon system:} velocity $\beta$, polar angle, and, for $\mathrm{BDT}(2\mu2\nu)$ only, the opening angle between the two muons;
    \item \textbf{angular observables:} $\Delta\eta$, $\Delta\phi$, and $\Delta R$ between the two muons, together with the $\mu^{-}$ helicity angle.
\end{itemize}
Variables strongly correlated with the dimuon invariant mass are deliberately excluded from the training to avoid sculpting the background mass distributions. 
A polygonal cut in the $\mathrm{BDT}(2\mu2\nu)$--$\mathrm{BDT}(4\mu)$ plane is then applied to maximise the signal significance. The effective production cross sections, preselection and BDT-cut efficiencies, and expected yields for the signal and background processes are summarised in Table~\ref{tab:h2mumu}. The expected yields quoted there correspond to the combined polygonal selection in the $\mathrm{BDT}(2\mu2\nu)$--$\mathrm{BDT}(4\mu)$ plane.

\begin{figure}[t!]
    \centering
    \includegraphics[width=\columnwidth]{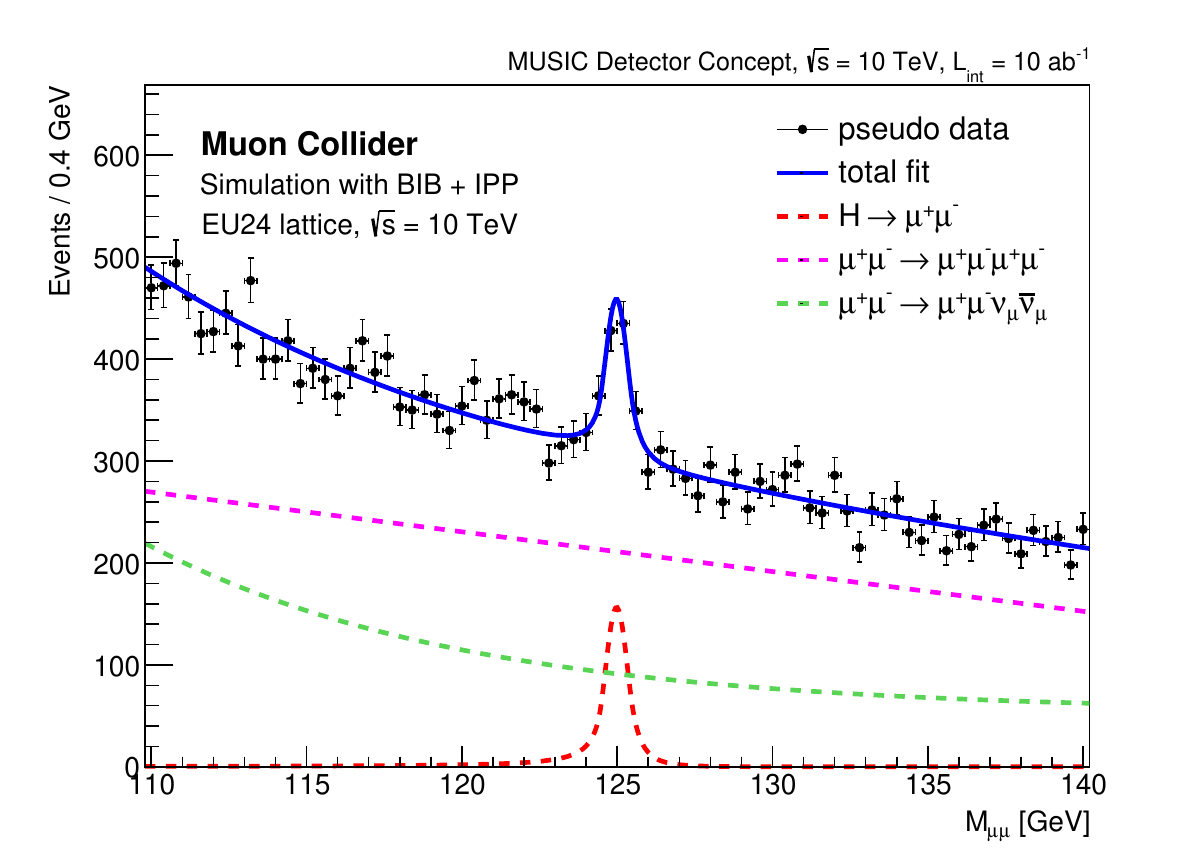}
    \caption{Dimuon invariant mass distribution for one pseudo-experiment (points), with the total fit (blue solid line) and its components: $H \to \mu^+\mu^-$ (red dashed line), $\mu^+\mu^-\mu^+\mu^-$ (magenta dotted line), and $\mu^+\mu^-\nu_\mu\bar{\nu}_\mu$ (green dotted line)
    \label{fig:hmumu_mass}}
\end{figure}

The statistical precision is determined from a fit to the dimuon invariant-mass distribution, with the signal yield treated as the parameter of interest. The model comprises three components: a double Crystal Ball function for the signal, a linear term for the $\mu^{+}\mu^{-}\mu^{+}\mu^{-}$ background, and an exponential-plus-constant term for the $\mu^{+}\mu^{-}\nu_{\mu}\bar{\nu}_{\mu}$ background. The free parameters are
the signal and background yields and the mean of the Crystal Ball function. 
An example fit is shown in Fig.~\ref{fig:hmumu_mass}. The expected precision is obtained from an unbinned extended maximum-likelihood fit performed for an ensemble of $10\,000$ pseudo-experiments, yielding a relative statistical precision on the $H \to \mu^{+}\mu^{-}$ cross section of
\begin{equation}
\frac{\Delta\sigma(H \to \mu^{+}\mu^{-})}{\sigma(H \to \mu^{+}\mu^{-})} = 9.7\% ,
\label{eq:Hmumu_prec}
\end{equation}
and a statistical uncertainty on the Higgs boson mass of $\Delta m_{H} = 51$~MeV.

\section{Higgs boson mass measurement}
\label{sec:mass}

The Higgs boson mass, $m_H$, is one of the fundamental parameters of the SM. Although the analyses presented here are not specifically optimised for its determination, they provide an estimate of the achievable uncertainty at a 10~TeV muon collider.

\begin{table}[t!]
    \centering
    \caption{Statistical uncertainty of the $m_H$ measurement obtained with different decay channels, assuming two experiments and an integrated luminosity of $\mathcal{L}_{\mathrm{int}}=10$ ab$^{-1}$ per experiment. The combination is also reported}
    \label{tab:Hm}
    \begin{tabular}{l|c}
    \toprule
    Decay channel & $\Delta m_H$ [MeV]\\
    \midrule
    $H \to b \bar{b}$ & 60 \\
    $H \to \gamma \gamma$ & 25 \\
    $H \to \mu^+ \mu^-$ & 36 \\
    \midrule
    Combination & 19 \\
    \bottomrule
    \end{tabular}
\end{table}

The mass can be measured by reconstructing the invariant mass of the Higgs boson decay products in channels with excellent mass resolution. In this study, the $H \to \gamma\gamma$ and $H \to \mu^+\mu^-$ channels are considered, since the excellent photon energy and muon momentum resolutions lead to narrow invariant-mass peaks above the continuum background, as discussed in Sects.~\ref{subsec:Htogammagamma} and \ref{subsec:Htomumu}. In addition, the $H \to b\bar{b}$ channel is also used for the $m_H$ measurement, since the shape of the dijet invariant-mass distribution depends directly on $m_H$. 
Given its large event yield, this channel further improves the overall precision of the mass measurement.

As explained in Sects.~\ref{subsec:Htobb}, \ref{subsec:Htogammagamma}, and \ref{subsec:Htomumu}, the Higgs boson yields in the $H\to b\bar{b}$, $H\to\gamma\gamma$, and $H\to\mu^+\mu^-$ channels are extracted through unbinned maximum-likelihood fits to the corresponding invariant-mass distributions. In each case, the signal is modelled by an analytical function with $m_H$ treated as a free parameter, and the statistical uncertainty on $m_H$ is determined from the fit.
The estimates are performed assuming an integrated luminosity of $\mathcal{L}_{\mathrm{int}}=10$ ab$^{-1}$ for a single experiment, and the resulting $\Delta m_H$ values are then extrapolated to the two-experiment scenario. The resulting statistical uncertainties are summarised in Table~\ref{tab:Hm}. The combined precision, obtained by assuming the three measurements to be statistically independent, is also reported.

The combined statistical uncertainty on $m_H$, found to be 19~MeV, is comparable to the precision expected at the end of the HL-LHC programme, where an uncertainty of approximately 21~MeV is anticipated. 
This result should be regarded as conservative, as it does not include all Higgs boson decay channels that can contribute to the mass determination. In particular, the $H\to ZZ^\ast\to\ell^+\ell^-\ell'^+\ell'^-$ channel, with electrons and muons in the final state, is expected to provide a highly precise measurement of $m_H$ due to its excellent invariant-mass resolution and clean experimental signature. This decay mode has not been included in the present study because the electron reconstruction algorithms are not yet optimised for precision analyses.
The inclusion of this and other complementary channels would further improve the overall precision. Consequently, the ultimate statistical uncertainty on $m_H$ at a 10~TeV muon collider is expected to be smaller than the value reported here.

\section{Double-Higgs production}
\label{sec:selfcoupling}

\begin{figure*}[!t]
    \centering
    \includegraphics[width=0.3\textwidth]{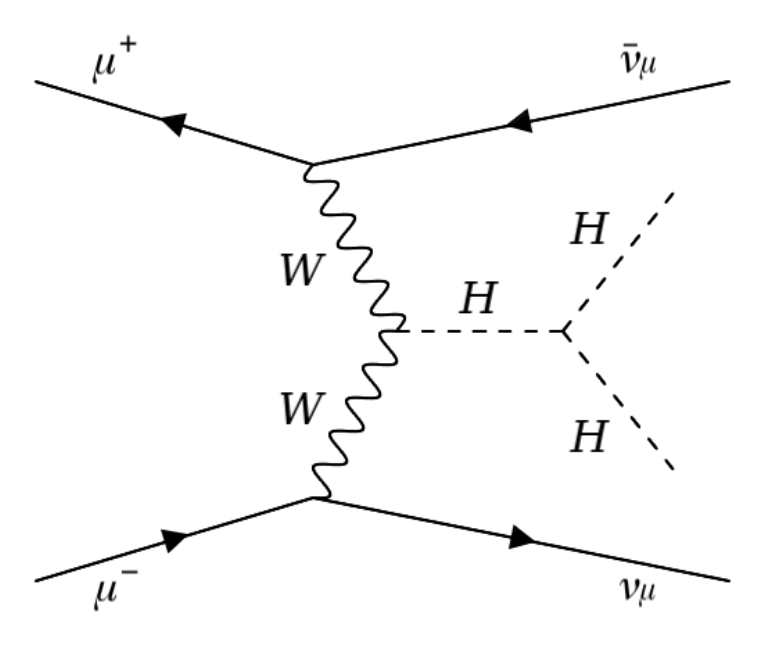} \hspace{0.4cm}
    \includegraphics[width=0.3\textwidth]{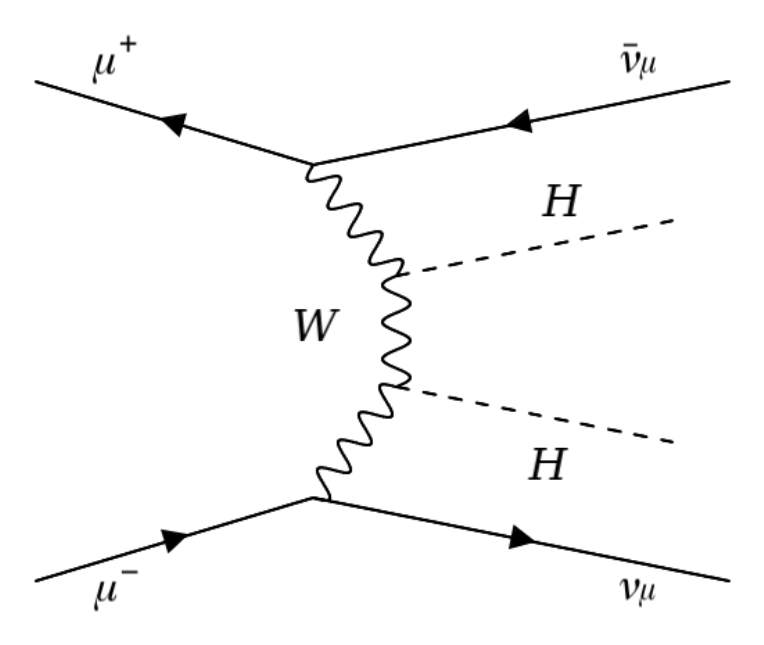} \hspace{0.4cm}
    \includegraphics[width=0.3\textwidth]{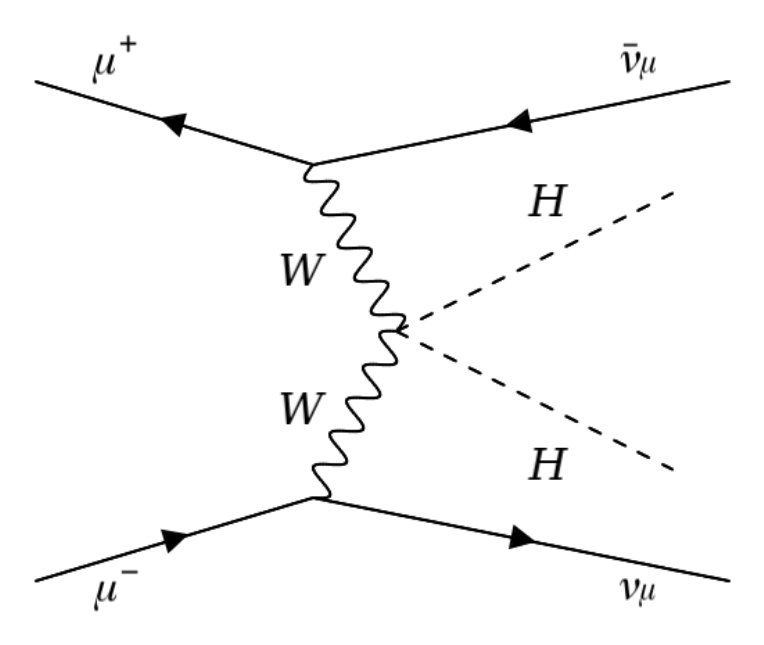}
    \caption{Tree-level Feynman diagrams contributing to double-Higgs production at a $\sqrt{s}=10$~TeV muon collider. The leftmost diagram contains the trilinear Higgs boson self-coupling vertex and is therefore directly sensitive to the Higgs self-coupling
\label{fig:HHFey}}
\end{figure*}

At a $\sqrt{s}=10$~TeV muon collider, the $H\!H$ production cross section is of the order of a few femtobarns, yielding the largest double-Higgs boson sample among the proposed future collider facilities over a relatively short data-taking period and with limited contributions from physics backgrounds. This provides the opportunity for a precise determination of the trilinear Higgs boson self-coupling and, consequently, a direct probe of the Higgs potential. Figure~\ref{fig:HHFey} shows the tree-level Feynman diagrams contributing to $H\!H$ production, with the leftmost diagram containing the trilinear Higgs boson self-coupling vertex.

\subsection{$H\!H\to b\bar{b}b\bar{b}$}
\label{sec:hh4b}

Double-Higgs production is studied in the $H\!H \to b\bar{b}b\bar{b}$ final state, exploiting the $H \to b\bar{b}$ decay for both Higgs bosons. 
This channel benefits from the large $H\to b\bar b$ branching fraction and provides the highest event yield among all $H\!H$ decay modes, making it the reference channel for this analysis.

Events are required to contain at least four reconstructed jets with $p_T^j > 20$~GeV, using the jet reconstruction procedure described in Ref.~\cite{music-paper}. 
The same $b$-tagging algorithm used in the single-$H\to b\bar{b}$ analysis of Sect.~\ref{subsec:Htobb} is applied here. The corresponding misidentification probabilities are parameterised as functions of the jet $p_T$ and $\eta$ and applied according to the flavour composition of the four-jet final state. At least two of the four selected jets are required to be $b$-tagged.
Two Higgs boson candidates are reconstructed from the four highest-energy jets by considering all possible pairings into two dijet systems, requiring each pair to contain at least one $b$-tagged jet. The combination that minimises
\[
F=\sqrt{(m_{12}-m_{H})^2+(m_{34}-m_{H})^2}\ ,
\label{F_figure_merit}
\]
is selected, where $m_{12}$ and $m_{34}$ are the invariant masses of the two dijet systems and $m_H$ is the nominal Higgs boson mass.

Two classes of physics background are considered. The largest contribution after event selection arises from single-Higgs production in association with a heavy-flavour quark pair, $\mu^+\mu^- \to H\,q_h\bar{q}_h\,(\nu_\ell\bar\nu_\ell,\,\ell\ell) \to$\linebreak $b\bar{b}\,q_h\bar{q}_h\,(\nu_\ell\bar\nu_\ell,\,\ell\ell)$,
with $q_h=b,c$, which does not proceed through the $H\!H\!H$ vertex. A comparable background arises from non-resonant four-heavy-flavour production accompanied by either a neutrino pair or two charged leptons,
$\mu^+\mu^- \to q_h\bar{q}_h q_h\bar{q}_h\,(\nu_\ell\bar\nu_\ell,\,\ell\ell)$.
The effective signal and background cross sections, selection efficiencies, including both preselection and $b$-tagging, and expected event yields for an integrated luminosity of 10 ab$^{-1}$ are summarised in Table~\ref{tab:HH}.

\begin{table}[b!]
    \centering
    \caption{Effective cross sections, total selection efficiencies, and expected events for signal and background processes in the $H\!H \rightarrow b\bar{b} b\bar{b}$ analysis with $\mu^+\mu^-$ collisions at 10 TeV assuming an integrated luminosity $\mathcal{L}_{\mathrm{int}}=10$ ab$^{-1}$. Here, $X$ indicates both $\nu_\ell\bar{\nu}_\ell$ and $\ell\ell$}
    \begin{tabular}{l|c c c c}
    \toprule
         Process & $\sigma_{\mathrm{eff}}$ [fb] &  $\epsilon_{\mathrm{sel}}$ [\%] &  $N_{\mathrm{exp}}$ \\
    \midrule
       $H\!H X \to b \bar{b} b \bar{b} X$         & $1.23$ & $18.5$ & $2276$ \\
    \midrule
       $H (\to b \bar{b}) q_{h}  \bar{q}_{h} X$  & $7.27$ & $15.6 $ & $11307$ \\
       $q_{h}  \bar{q}_{h} q_{h}  \bar{q}_{h} X$ & $10.9$ & $9.0$ & $9787$ \\
    \bottomrule
    \end{tabular}
    \label{tab:HH}
\end{table}

\begin{figure}[!t]
    \centering
    \includegraphics[width=\columnwidth]{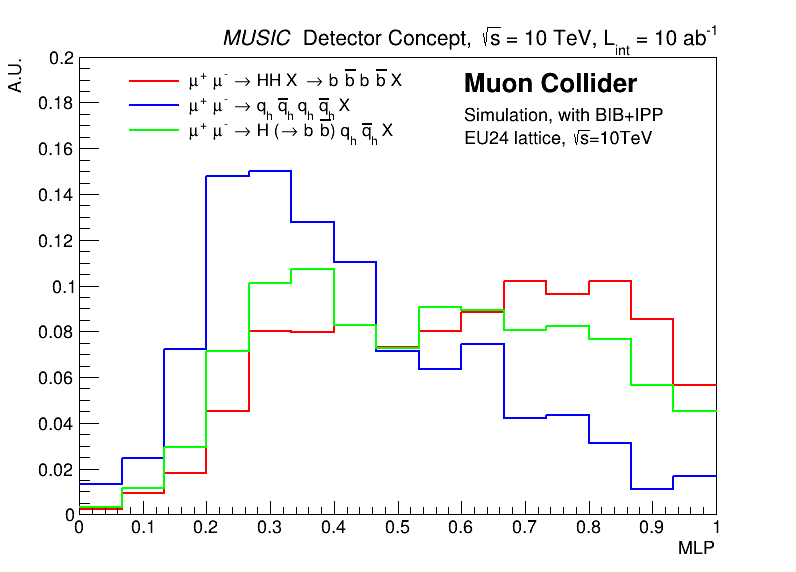}
    \caption{Distributions of the MLP output for the $H\!H$ signal (red solid line) and the main background contributions (blue and green solid lines). The distributions are normalised to unit area
    \label{fig:MLP_HH}}
\end{figure}

Separation between the signal and the backgrounds is performed using a Multilayer Perceptron (MLP)~\cite{TMVA} trained on the following kinematic observables:
\begin{itemize}
    \item the invariant masses of the leading and subleading Higgs boson candidates, defined as the dijet systems with the higher and lower total transverse momentum, respectively;
    \item the opening angle between the two Higgs boson candidates;
    \item the polar angle of the highest-$p_T$ jet in each Higgs boson candidate;
    \item the transverse momenta of the four selected jets.
\end{itemize}
The resulting MLP output distributions for the signal and background processes are shown in Fig.~\ref{fig:MLP_HH}, demonstrating good discrimination among the different event classes.

\begin{table*}[b!]
    \centering
    \caption{Effective production cross sections, selection efficiencies, and expected event yields for signal and background processes in the $H\!H \rightarrow b\bar{b}WW^\ast$ analysis with $\mu^+\mu^-$ collisions at 10 TeV and $\mathcal{L}_{\mathrm{int}}=10$ ab$^{-1}$. For each process, the values are reported for the electron and muon channel analyses. Due to the presence of misidentified leptons, $H\!H$ events containing an electron contribute to the background in the muon channel, and vice versa}
    \label{tab:hh2bbww}
    \begin{tabular}{l c | c c c}
    \toprule
    Process & Channel & $\sigma_{\text{eff}}$ [fb] & $\epsilon_{\text{sel}}$ [\%] & $N_{\text{exp}}$ \\
    \midrule
    \multirow{2}{*}{$H\!H \rightarrow b\bar{b}WW^\ast\rightarrow b\bar{b}q\bar{q}^\prime\mu\nu_\mu$} & $\mu$ & \multirow{2}{*}{0.136} & 19.3 & 262 \\
                                                                    & $e$   &                        & 6.9  & 94  \\
    \addlinespace
    \multirow{2}{*}{$H\!H \rightarrow b\bar{b}WW^\ast\rightarrow b\bar{b}q\bar{q}^\prime e\nu_e$} & $\mu$ & \multirow{2}{*}{0.136} & 3.1  & 42  \\
                                                                & $e$   &                        & 20.0 & 272 \\
    \midrule
    \multirow{2}{*}{$q\bar{q}H$} & $\mu$ & \multirow{2}{*}{27.6} & 3.2  & $8.75\times10^{3}$ \\
                           & $e$   &                       & 10.0 & $2.76\times10^{4}$ \\
    \addlinespace
    \multirow{2}{*}{$q\bar{q}q_h\bar{q}_h \; (q_h=b,c)$} & $\mu$ & \multirow{2}{*}{253} & 1.9 & $4.74\times10^{4}$ \\
                                             & $e$   &                      & 5.5 & $1.40\times10^{5}$ \\
    \bottomrule
    \end{tabular}
\end{table*}

The expected statistical precision on the signal yield is evaluated using pseudo-datasets generated from the expected signal and background event yields. The MLP output distribution in each pseudo-dataset is fitted using the same procedure adopted for the single-Higgs analyses. The distribution of the fitted signal yields is then used to determine the expected statistical uncertainty, corresponding to a relative uncertainty on the production cross section of
\begin{equation}
    \frac{\Delta \sigma (H\!H \rightarrow b\bar{b} b\bar{b})}{\sigma (H\!H \rightarrow b\bar{b} b\bar{b})} = 6.0\%\ .
\end{equation}

\subsection{$H\!H \rightarrow b\bar{b}WW^\ast$}
\label{subsec:hhbbww}

The $H\!H\to b\bar{b}WW^\ast$ channel provides a complementary probe of double-Higgs production. Although its branching fraction is smaller than that of the $H\!H\to b\bar{b}b\bar{b}$ channel, the presence of an isolated charged lepton significantly reduces the background, resulting in a favourable signal-to-background ratio.
The $H\!H\to b\bar{b}WW^\ast$ process is studied in the semileptonic final state, where one Higgs boson decays into $b\bar{b}$ and the other into $WW^\ast\to q\bar{q}^\prime\ell\nu_\ell$ ($\ell=e,\mu$) with a branching fraction of approximately 3.5\% for each lepton flavour. 
Separate analyses are performed for the electron and muon channels. 

Each analysis proceeds in three stages.
First, a loose preselection is applied to select $H\!H\to b\bar{b}WW^\ast$ candidate events and suppress the dominant background processes. Events are required to contain at least four reconstructed jets with $p_T^{j} > 10$~GeV, of which at least one is identified as a $b$ jet. Additionally, at least one reconstructed charged lepton ($e$ or $\mu$ depending on the analysis) with $p_T^\ell > 5$~GeV is required. The preselection efficiencies and expected yields for the signals and the most significant background processes are reported in Table~\ref{tab:hh2bbww}.

Second, the $H\to b\bar{b}$, hadronic $W$, and $H\to WW^\ast$ candidates are reconstructed as follows:
\begin{itemize}
\item The $H\to b\bar{b}$ candidate is reconstructed by pairing the $b$-tagged jets. If exactly two $b$-tagged jets are present, they are assigned to the Higgs boson candidate. If only one $b$-tagged jet is identified, it is combined with the untagged jet yielding an invariant mass closest to $m_H$. If more than two $b$-tagged jets are present, the pair with invariant mass closest to $m_H$ is selected.
\item The hadronic $W$ boson candidate is reconstructed from a non-tagged jet pair with the invariant mass closest to $m_W$.
\item The $H\to WW^\ast$ candidate is built by combining the hadronic $W$ candidate with the highest-$p_T$ lepton.
\end{itemize}

\begin{figure*}[t]
    \centering
    \includegraphics[width=0.49\linewidth]{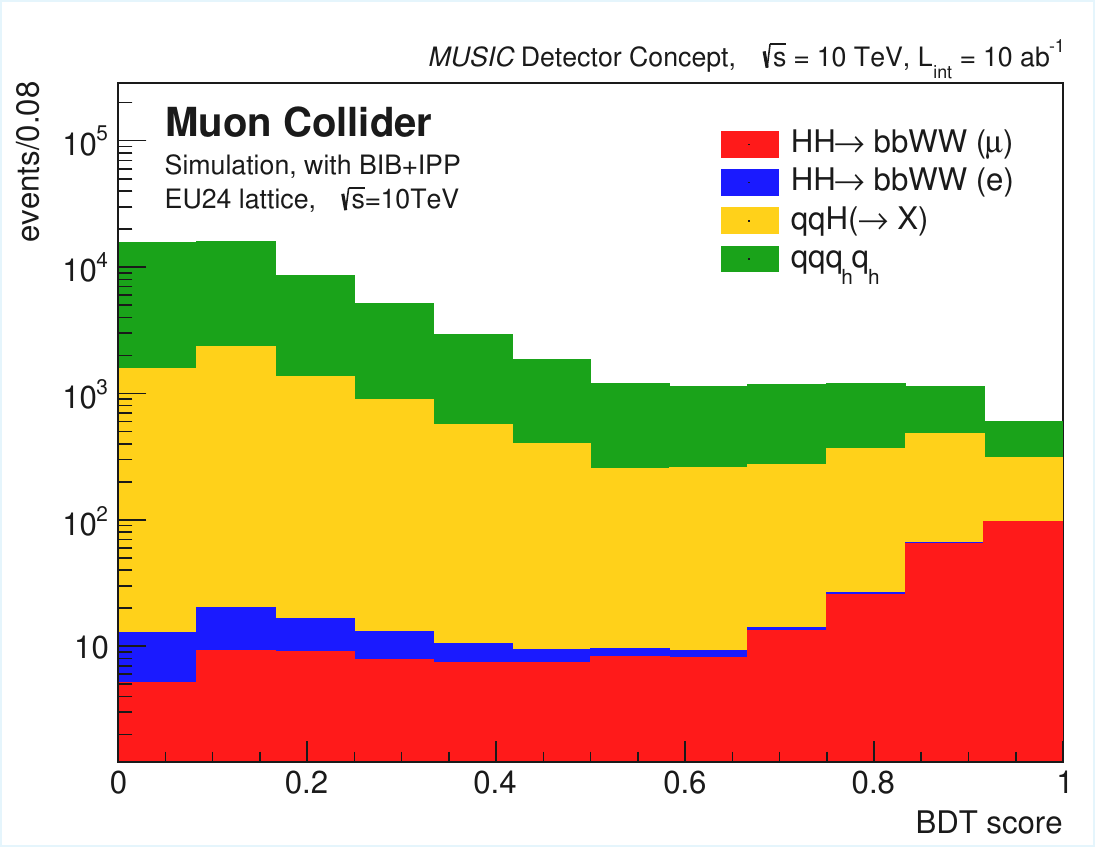}
    \hfill
    \includegraphics[width=0.49\linewidth]{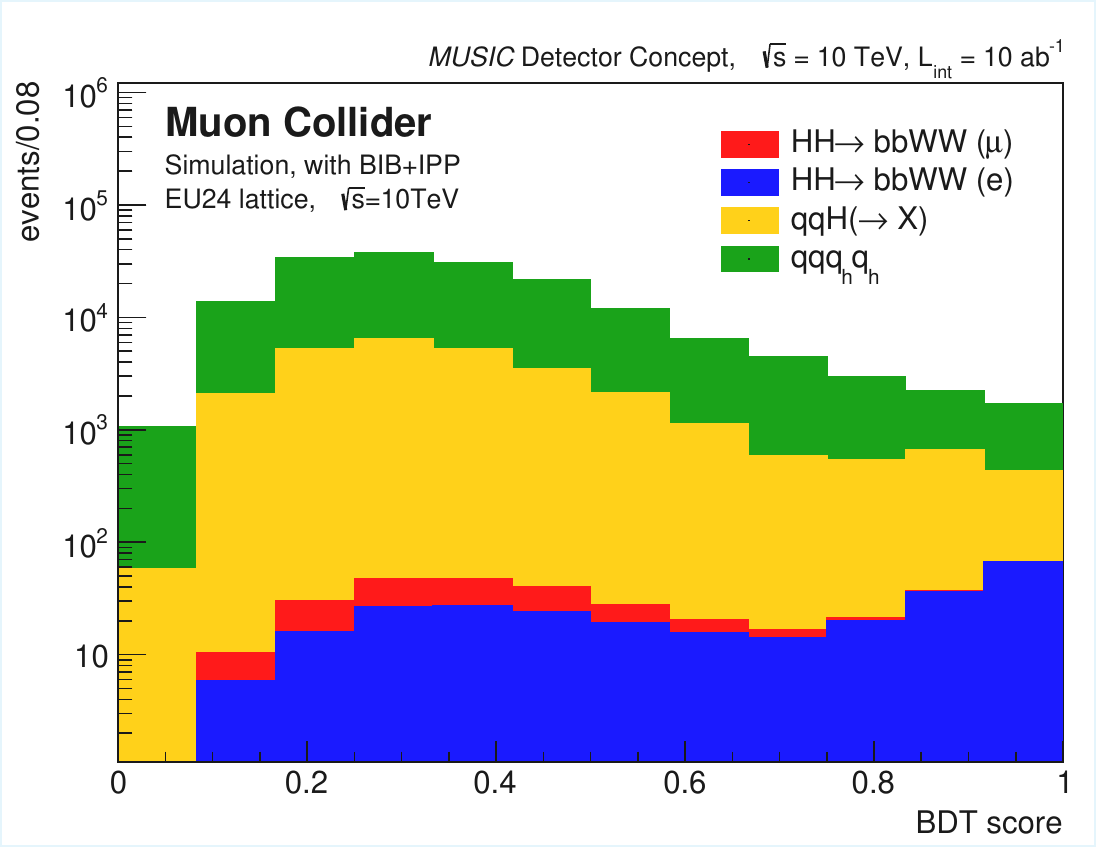}
    \caption{Stacked BDT score distributions of the signal and background processes for the muon-channel (left) and electron-channel (right) analyses
    \label{fig:hh2bbww}}
\end{figure*}

Third, a BDT~\cite{XGBoost} is employed to improve the separation between the signal and background processes. The BDT takes the following variables as input:
\begin{itemize}
    \item \textbf{lepton:} transverse momentum, pseudorapidity, isolation, transverse impact parameter, and, for the electron channel, the associated calorimeter energy;
    \item \textbf{event kinematics:} the scalar sum of the transverse momenta of the selected jets and lepton, the total invariant mass of the selected jets and lepton, and the invariant mass of the $H\to b\bar{b}$ candidate;   
    \item \textbf{angular observables:} the opening angles between the lepton and the hadronic $W$ boson and between the lepton and the $H\to WW^\ast$ candidate, both evaluated in the laboratory frame, and the opening angle between the two Higgs boson candidates in the $H\!H$ centre-of-mass frame.
\end{itemize}
Lepton isolation $I_\ell$ is defined with respect to the nearest jet as:
\[
I_\ell =
\begin{cases}
2 & \text{if } \Delta R(j,\ell) > 0.5\ ,\\
p_T^\ell/p_T^{j} & \text{if } \Delta R(j,\ell) \leq 0.5\ .
\end{cases}
\]
The stacked BDT score distributions of the signal and background processes are shown in Fig.~\ref{fig:hh2bbww} for the muon (left) and electron (right) channel analyses.

In both channels, the statistical precision of the signal yield is evaluated using a toy Monte Carlo study, based on the fit to the BDT output distribution. An extended binned maximum-likelihood model is constructed with four components: $H\!H \rightarrow b\bar{b}WW^\ast\,(\to q\bar{q}^\prime\mu\nu_\mu)$, $H\!H \rightarrow b\bar{b}WW^\ast\,(\to q\bar{q}^\prime e\nu_e)$, the inclusive single-Higgs background $q\bar{q}H$, and the four-jet background $q\bar{q}q_h\bar{q}_h$. The $q\bar{q}H$ and $q\bar{q}q_h\bar{q}_h$ yields are fixed in the fit, since their cross sections are predicted with a relative uncertainty much smaller than the expected sensitivity to the $H\!H$ signal, so that their normalisation has negligible impact on the result. The signal yields are treated as free parameters of the fit.

The expected statistical precision on the signal yield is evaluated independently in the electron and muon channels, yielding 53\% and 22\%, respectively.
The two sensitivity estimates are then combined assuming statistical independence. This assumption is justified at the simulation level because the selected event samples are disjoint and the corresponding background samples are generated independently. Correlations arising from common systematic uncertainties will need to be taken into account once collider data become available.

The combined expected relative statistical uncertainty on the $H\!H\to b\bar{b}WW^\ast$ production cross section is
\begin{equation}
    \frac{\Delta \sigma(H\!H \rightarrow b\bar{b}WW^\ast)}{\sigma (H\!H \rightarrow b\bar{b}WW^\ast)} = 20\%\ .
\end{equation}
The statistical precision in the electron channel is significantly worse than in the muon channel. This is primarily due to the more challenging reconstruction and identification of electrons in multijet events, which increases the probability of selecting an incorrect electron candidate and reduces the separation between signal and background.
Improvements in electron reconstruction and identification, which are beyond the scope of the present study, are therefore expected to enhance the sensitivity of the $H\!H\to b\bar{b}WW^\ast$ analysis. At present, the combined measurement is largely driven by the muon channel.

\section{Summary of Higgs production cross-section sensitivities}
\label{sec:xsec_summary}

\begin{table}[b!]
    \centering
    \caption{Expected statistical sensitivities on the production cross sections for the benchmark single- and double-Higgs channels considered in this study. The sensitivities correspond to the combination of two statistically independent experiments, each collecting an integrated luminosity of 10~ab$^{-1}$, and are obtained by scaling the event yields from the single-experiment scenario}
    \label{tab:xsec_summary}
    \begin{tabular}{l|c}
    \toprule
    Channel & $\Delta\sigma/\sigma$ [\%] \\
    \midrule    
    $H\to b\bar{b}$                                                  &  0.18 \\
    $H \to WW^{*} \to q\bar{q}^\prime\ell\nu_\ell$ $(\ell = \mu, e)$ &  0.35 \\
    $H \to ZZ^{*} \to q\bar{q}\mu^+\mu^-$                            &  4.2  \\
    $H \to \gamma\gamma$                                             &  2.6  \\
    $H\to \mu^+\mu^-$                                                &  6.9  \\
    \midrule    
    $H\!H \to b\bar{b}b\bar{b}$                                      & 4.2 \\
    $H\!H \to b\bar{b}WW^\ast \to b\bar{b} \, q\bar{q}^\prime\ell\nu_\ell$ $(\ell = \mu, e)$ & 14 \\   
    \bottomrule
    \end{tabular}
\end{table}

The results presented in the previous sections are based on a single experiment collecting an integrated luminosity of 10~ab$^{-1}$. The baseline IMCC muon collider design foresees two interaction points, each instrumented with a general-purpose detector and capable of operating simultaneously without any reduction in the delivered luminosity. The expected statistical sensitivities are therefore extrapolated to the nominal two\hyp{experiment} configuration, corresponding to a total integrated luminosity of 20~ab$^{-1}$.
The results are summarised in Table~\ref{tab:xsec_summary}. 

For the most abundant single-Higgs decay channels, sub-percent precision is achieved for the measurements of the $H\to b\bar{b}$ and $H\to WW^\ast$ production cross sections, reaching relative uncertainties of 0.18\% and 0.35\%, respectively. For these channels, experimental systematic uncertainties are expected to become comparable to, or even dominate over, the statistical uncertainty. Their evaluation will require collider data and dedicated calibration procedures and is therefore beyond the scope of the present study. To allow a consistent comparison with the projected performance of other future collider facilities, only statistical uncertainties are reported in Table~\ref{tab:xsec_summary}.

The rare decay channels $H\to\gamma\gamma$, $H\to ZZ^\ast$, and $H\to\mu^+\mu^-$ are expected to be measured with statistical precisions of 2.6\%, 4.2\%, and 6.9\%, respectively. 
For double-Higgs production, the $H\!H\to b\bar{b}b\bar{b}$ channel reaches a statistical precision of 4.2\%, while the semileptonic $H\!H\to b\bar{b}WW^\ast$ channel is measured with a precision of about 14\%. These results are competitive with, and in several cases exceed, the projected sensitivities of other proposed future collider facilities~\cite{ESPPUphysics}, highlighting the unique potential of a 10~TeV muon collider for precision studies of the Higgs self-interactions.

\section{Trilinear Higgs self-coupling}
\label{sec:trilinear}

The mechanism of electroweak symmetry breaking~\cite{higgs1,higgs2,higgs3,higgs4} is described in the SM by the Higgs scalar potential $V\!(h)$. After EWSB, it can be written as
\[
    V\!(h)=\frac{1}{2} m_H^2 h^2+\lambda_3 vh^3+\frac{1}{4} \lambda_4 h^4\ , 
    \label{eq:higgspotential}
\]
where $h$ denotes the physical Higgs boson field remaining after EWSB. The Higgs boson mass is measured to be $m_H = 125.13 \pm 0.11$~GeV~\cite{PDG}, while $v=1/\sqrt{\sqrt{2}\, G_F} \simeq 246$~GeV is the vacuum expectation value of the Higgs field, with $G_F$ the Fermi constant. In the SM, the trilinear and quartic Higgs self-couplings are uniquely determined by the Higgs potential, $\lambda_3=\lambda_4=\lambda=m_H^2/(2v^2)$.
A direct measurement of the Higgs self-couplings therefore provides a unique experimental probe of the Higgs potential and of the mechanism of electroweak symmetry breaking. Any deviation from the SM prediction would constitute clear evidence for physics beyond the SM.

\begin{figure*}[t]
    \centering
    \includegraphics[width=0.49\linewidth]{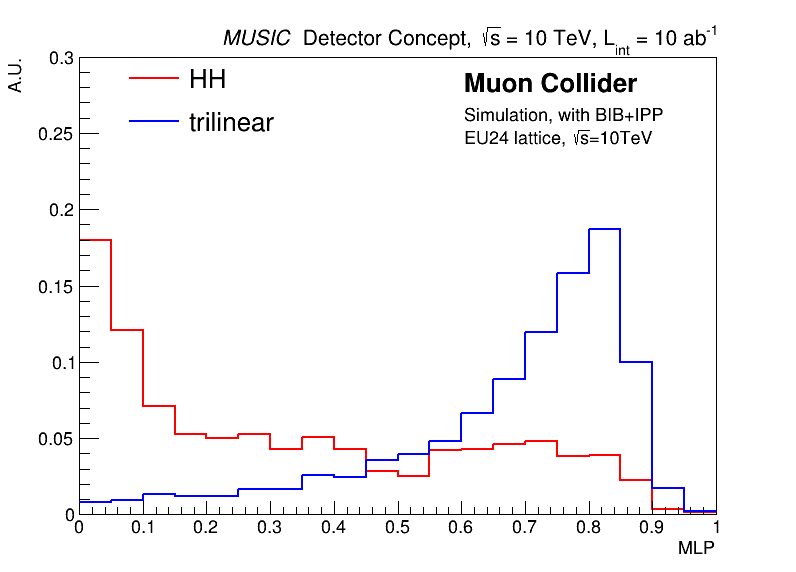}
    \hfill
    \includegraphics[width=0.49\linewidth]{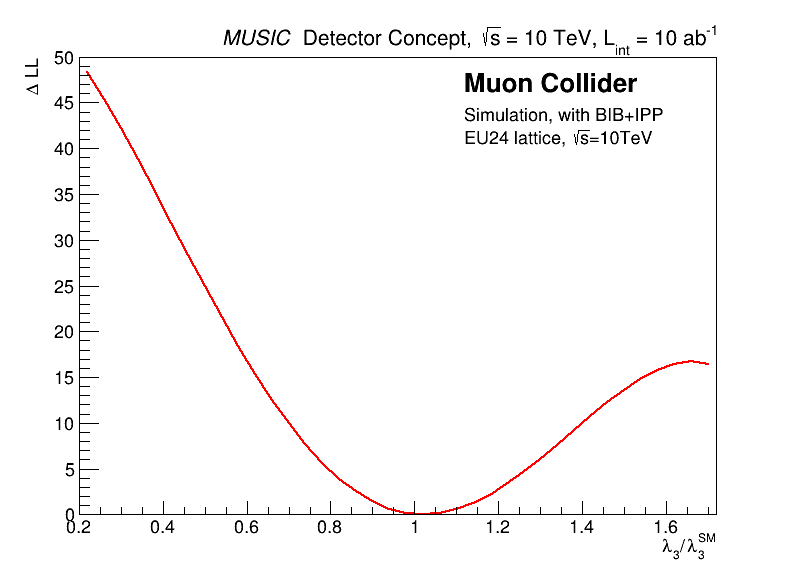}
    \caption{In the left panel, output distributions for the MLP applied to distinguish the $H\!H$ events sensitive to $\lambda_3$ from the other $H\!H$ contributions. On the right, $\Delta LL$ as a function of the $\kappa_3$ hypothesis
    \label{fig:DLL}}
\end{figure*}

A measurement of the trilinear self-coupling $\lambda_3$ requires isolating the contribution of the double-Higgs production mechanism involving an intermediate off-shell Higgs boson, $H^\ast$, shown as the leftmost diagram in Fig.~\ref{fig:HHFey}. This contribution exhibits kinematic properties that differ from those of the remaining production mechanisms, allowing its fraction to be statistically constrained. The analysis exploits these differences following the strategy adopted in Ref.~\cite{Higgs3TEV} and proceeds as follows:
\begin{itemize}
    \item Eleven double-Higgs MC samples are generated with \textsc{Whizard} for values of the coupling modifier $\kappa_3=\lambda_3/\lambda_{3}^{\mathrm{SM}}$ ranging from 0.2 to 1.8. The generated events are processed through the detector simulation and reconstruction chain. The sample with $\kappa_3=1$ corresponds to the SM signal hypothesis.

    \item Two MLPs are trained independently to separate the signal from the backgrounds and to isolate the $\lambda_3$-sensitive events from the remaining $H\!H$ signal sample:
    \begin{enumerate}
        \item the first MLP is the same classifier used in the $H\!H \to b\bar{b}b\bar{b}$ cross-section analysis of Sect.~\ref{sec:hh4b}, trained to separate the SM signal ($\kappa_3=1$) from the physics backgrounds;
        \item the second MLP is trained on four kinematic variables to discriminate the contribution involving the off-shell Higgs boson from the remaining $H\!H$ production mechanisms. The most discriminating variables are the opening angle between the two Higgs bosons in the laboratory frame, the polar angles of the highest-$p_T$ jet in each Higgs boson candidate, and the helicity angle of the two Higgs boson candidates. The resulting MLP output distributions for the two event classes are shown in Fig.~\ref{fig:DLL} (left).
    \end{enumerate}

    \item For each value of $\kappa_3$, signal and background two-di\-men\-sion\-al templates are built from the joint distribution of the two MLP outputs, normalised to the corresponding expected event yields.

    \item Pseudo-datasets are generated using the two-di\-men\-sion\-al template for the $\kappa_3=1$ hypothesis. For each pseudo-experiment, the log-likelihood difference\linebreak $\Delta LL=-\Delta\log(L)$ is computed as a function of $\kappa_3$ by comparing the pseudo-data distribution with the templates built for each $\kappa_3$ hypothesis.

    \item The resulting log-likelihood profile is fitted with a fourth-degree polynomial. The 68\% confidence interval on $\kappa_3$ is defined by the region where the fitted curve satisfies $\Delta LL<0.5$ around the Standard Model value, $\kappa_3=1$.
\end{itemize}

The likelihood scan obtained with this procedure is shown in Fig.~\ref{fig:DLL} (right). The resulting constraint on the  trilinear Higgs boson self-coupling is
\begin{equation}
    0.94 < \kappa_3 < 1.08 \qquad (68\%~\mathrm{CL})\ . 
    \label{eq:HH_k3}
\end{equation}
This corresponds to an expected precision of approximately 7\% on $\kappa_3$. 
The $H\!H\to b\bar{b}WW^\ast$ channel is not included in the extraction of $\kappa_3$. Its expected statistical precision on the $H\!H$ production cross section, approximately $20\%$, is considerably poorer than the $6\%$ precision obtained in the $H\!H\to b\bar{b}b\bar{b}$ channel and therefore does not provide a significant improvement in the combined constraint.

The result quoted in Eq.~(\ref{eq:HH_k3}) corresponds to a single experiment collecting an integrated luminosity of $10$~ab$^{-1}$. 
Assuming two statistically independent experiments with equivalent performance, the expected 68\%~CL interval on $\kappa_3$ is
\begin{equation}
    0.97 < \kappa_3 < 1.05 \qquad (68\%~\mathrm{CL})\ . 
    \label{eq:HH_k3_20ab}
\end{equation}
Based on current projections, this would represent the most precise determination of the trilinear Higgs boson self-coupling among all proposed future collider facilities, achievable after approximately five years of operation.

\section{Conclusions and outlook}
\label{sec:conclusions}

This review has presented projected measurement precisions within the Higgs physics programme of a 10~TeV muon collider, including precision determinations of\linebreak Higgs boson production cross sections, the Higgs boson mass, and the trilinear self-coupling. The studies are based on detailed simulations of the MUSIC detector, including the effects of machine-induced background, and demonstrate the strong potential of a high-energy muon collider for precision Higgs physics. With an integrated luminosity of 10~ab$^{-1}$ per experiment, corresponding to approximately five years of operation at the design luminosity, such a facility could achieve competitive precision on a substantially shorter timescale than other proposed future colliders.

For single-Higgs production, the large vector-boson-fusion cross section combined with the baseline integrated luminosity of 20~ab$^{-1}$ enables unprecedented statistical precision over a broad range of Higgs boson decay channels. Sub-percent statistical precision is expected for the dominant $H\to b\bar{b}$ and $H\to WW^\ast$ production cross sections, while percent-level sensitivities are achieved for the rare decays $H\to\gamma\gamma$, $H\to ZZ^\ast$, and $H\to\mu^+\mu^-$. The Higgs boson mass is expected to be measured with a statistical precision of about 19~MeV using only the $H\to b\bar{b}$, $H\to\gamma\gamma$, and $H\to\mu^+\mu^-$ channels. Since this study does not include all experimentally accessible decay modes, in particular the clean $H\to ZZ^\ast\to4\ell$ final state, this estimate should be regarded as conservative.

The large double-Higgs production rate at multi-TeV energies provides unique sensitivity to the Higgs boson self-interactions. Measurements in the $H\!H\to b\bar b b\bar b$ and $H\!H\to b\bar bWW^\ast$ final states are expected to determine the double-Higgs production cross section with relative statistical uncertainties of approximately 4\% and 14\%, respectively. Using the $H\!H\to b\bar{b}b\bar{b}$ channel alone, the trilinear Higgs self-coupling can be determined with an expected precision of about 7\% for one experiment. Extrapolating to the baseline configuration with two interaction points yields an expected precision of approximately 5\%, placing a multi-TeV muon collider among the most powerful proposed facilities for directly probing the Higgs potential.

The studies presented in this review should be regarded as a first assessment of the Higgs physics capabilities of a 10~TeV muon collider rather than the ultimate performance. Several analyses can benefit from further developments in reconstruction algorithms, particularly for electrons and forward jets, where machine-induced background poses significant challenges. Additional Higgs boson decay channels, improved flavour-tagging techniques, and more sophisticated multivariate analyses are expected to further enhance the overall sensitivity.

Beyond the trilinear self-coupling, the sizeable triple-Higgs production cross section at multi-TeV energies opens the possibility of directly probing the quartic Higgs self-coupling as discussed in Ref.~\cite{Chiesa}. 
A quantitative  assessment of the sensitivity to the triple-Higgs production cross section and to the quartic self-coupling requires a dedicated detector-level analysis, including the evaluation of the relevant SM backgrounds and the development of jet reconstruction algorithms optimised for the forward region  where most of the signal is produced. Such a study could represent a natural continuation of the present work.

Overall, the results presented in this review demonstrate that a 10~TeV muon collider would provide an exceptionally powerful laboratory for precision Higgs physics. It combines an unprecedented single- and double\hyp{Higgs} production programme with unique access to Higgs boson self-interactions, offering the opportunity to explore the structure of the Higgs sector with a precision unmatched by existing facilities and highly competitive with that of any proposed future collider.

\begin{acknowledgements}
We are grateful to the International Muon Collider Collaboration for its support.
We acknowledge funding from the Italian National Institute for Nuclear Physics (INFN), the University of Padua, and the European Organization for Nuclear Research (CERN).
We also acknowledge CloudVeneto for providing computing and storage resources.
This work was supported by the European Union's Horizon Europe Research and Innovation Programme through the Research Infrastructures INFRADEV Grant Agreement No.~101094300.
\end{acknowledgements}

\bibliographystyle{spphys}        
\bibliography{bibliography}    

%

\end{document}